\documentclass[12pt]{article}
\pdfoutput=1
\usepackage{jheppub, amsmath,amssymb,amsfonts,amsxtra, mathrsfs, makeidx,graphics,graphicx,amsthm,epsfig, ytableau,bm,longtable,float, color,tikz,mathtools}
\usetikzlibrary{positioning,matrix,arrows}
\usepackage{bbm}

\newcommand{\ba}{\begin{array}}
\newcommand{\ea}{\end{array}}
\newcommand{\bi}{\begin{itemize}}
\newcommand{\ei}{\end{itemize}}
\def\vec#1{\bm{#1}}
\def\bea#1\eea{\allowdisplaybreaks \begin{align}#1\end{align}}
 \newcommand{\ben}{\begin{enumerate}}
\newcommand{\een}{\end{enumerate}}
\newcommand{\bean}{\begin{eqnarray*}}
\newcommand{\eean}{\end{eqnarray*}}
\newcommand{\eref}[1]{(\ref{#1})}

\newcommand{\tref}[1]{Table~\ref{#1}}

\newcommand{\nn}{\nonumber}

\newcommand{\tr}{\mathrm{Tr}}
\newcommand{\PE}{\mathop{\rm PE}}

\newcommand{\BC}{\mathbb{C}}

\newcommand{\BQ}{\mathbb{Q}}

\newcommand{\BZ}{\mathbb{Z}}
\newcommand{\BF}{\mathbb{F}}

\newcommand{\CP}{\mathbb{CP}}

\newcommand{\comment}[1]{}

\newcommand{\CM}{{\cal M}}

\newcommand{\CN}{{\cal N}}

\newcommand{\CH}{{\cal H}}

\newcommand{\CV}{{\cal V}}

\newcommand{\ie}{{\it i.e.}}
\newcommand{\eg}{{\it e.g.}}

\newcommand{\ud}{\mathrm{d}}

\newcommand{\fflat}{\mathcal{F}^\flat}

\newcommand{\rk}{\mathrm{rk}}

\title{{\Large The ADHM-like Constructions for Instantons on $\mathbb{CP}^2$ and Three Dimensional Gauge Theories}}
\author[a]{Noppadol Mekareeya}
\author[b]{and Diego Rodr\'iguez-G\'omez}

\affiliation[a]{
Theory Group, Physics Department, CERN, \\CH-1211, Geneva 23, Switzerland
}
\affiliation[b]{
Department of Physics, Universidad de Oviedo, \\
Avda. Calvo Sotelo 18, 33007, Oviedo, Spain
}

\emailAdd{noppadol.mekareeya@cern.ch,d.rodriguez.gomez@uniovi.es}

\preprint{
{\small
\begin{flushright}
CERN-PH-TH/2014-057\\
NSF-KITP-14-039
\end{flushright}
}
}

\abstract{We study the moduli spaces of self-dual instantons on $\CP^2$ in a simple group $G$.  When $G$ is a classical group, these instanton solutions can be realised using ADHM-like constructions which can be naturally embedded into certain three dimensional quiver gauge theories with 4 supercharges. The topological data for such instanton bundles and their relations to the quiver gauge theories are described.  Based on such gauge theory constructions, we compute the Hilbert series of the moduli spaces of instantons that correspond to various configurations.  The results turn out to be equal to the Hilbert series of their counterparts on $\mathbb{C}^2$ upon an appropriate mapping, in agreement with the result of [arXiv:0802.3120]. We check the former against the Hilbert series derived from the blowup formula for the Hirzebruch surface $\BF_1$ and find an agreement. The connection between the moduli spaces of instantons on such two spaces is explained in detail.
}

\begin{document}
\setcounter{tocdepth}{2}
\maketitle

\section{Introduction}

Understanding the dynamics of gauge theories is a central question in modern theoretical physics. In particular, many aspects involving non-perturbative effects remain yet to be fully understood. Over the recent past it has become clear that placing the gauge theory under study on different manifolds can help to unravel its properties, which may become more manifest upon considering non-trivial spacetimes. In this paper we take a step along these directions by considering pure gauge theories on $\CP^2$, focusing in particular on their instanton sector.

Since $\CP^2$ is a topologically non-trivial complex manifold, and hence has a preferred orientation, there is an intrinsic difference between self-dual and anti-self-dual instantons on it. More precisely, the key issue is wether the connections to describe have the same or the opposite duality properties as the K\"ahler form, which, upon choosing the preferred orientation induced by the complex structure, is taken to be self-dual. In the following we will concentrate on self-dual instantons. As we will review below, in the case of $\CP^2$ these are the connections for which an ADHM-like construction exists \cite{buchdal1,buchdal2,buchdal3,king, bryansanders2}. Note that this is contrary to the standard $S^4$ case, where ADHM \cite{ADHM} (see also \cite{Atiyah:1979iu, DonaldsonKronheimer}) constructs anti-self-dual connections. This is because in the case at hand there is an extra twist, as the ADHM-like construction really gives anti-self-dual connections on $\overline{\CP}^2$, which, upon orientation reversal, become self-dual connections on $\CP^2$.

The structure of the self-dual instanton sector has interesting topological properties. These are very reminiscent of the properties of other topological ``excitations" in gauge theory, such as line operators and surface operators (see \textit{e.g.} \cite{Aharony:2013hda} and references therein). In fact, due to the non-trivial topology of $\CP^2$, our instantons are characterized not only by their instanton number associated to an element of $H^4(\CP^2)$, but also by other characteristic classes depending on the gauge group. For example, as we will see below, instantons in the unitary group can be naturally viewed as instantons in the group $SU(N)/\mathbb{Z}_N$. Therefore, they come with a certain $N$-ality which suggests that the would-be first Chern number is really a mod $N$ integer.

Since $S^5$ can be regarded as the $S^1$ Hopf fibering over $\CP^2$, it is natural to suspect that our instantons might be relevant to the study of 5d gauge theories on $S^5$. However, as raised above, the instantons which we will discuss have the same duality properties as the K\"ahler form. Hence our instantons are different from the ones appearing, at least naively, in the computation of the partition function of a five dimensional gauge theory on $S^5$  \cite{Kim:2012qf} or the partition function of a four dimensional gauge theory on $\CP^2$. However, \textit{via} the blowup formula, our instanton partition functions do receive contributions from two fixed points, each of which contains the partition function of instantons on $\mathbb{C}^2$. The physical significance of this fact, as well as other possible implications for supersymmetric partition function computations, remain yet to be fully explored.

In the case of instantons on $\mathbb{C}^2$ (or its conformal compactification $S^4$), the ADHM construction allows to find all anti-self-dual connections. As it is well-known, it can be embedded into the Higgs branch of a gauge theory with 8 supercharges, therefore free of quantum corrections. In turn, such theory arises naturally in string theory as the worldvolume theory of a D$p$-D$(p+4)$ system \cite{Douglas:1995bn, Witten:1995gx, Douglas:1996sw}. Indeed, string theory provides a nice perspective on ADHM, as from the point of view of the D$(p+4)$ brane dissolving D$p$ branes is done by turning on a worldvolume instanton on the 4 transverse coordinates to the D$p$ inside the D$(p+4)$. In turn, the ADHM construction arises naturally as the Higgs vacua of the D$p$-D$(p+4)$ theory describing microscopically the system. In the past this construction has been used to compute the Hilbert series of the instanton moduli space \cite{Benvenuti:2010pq,Hanany:2012dm} which can be interpreted as the Nekrasov instanton partition function.  In fact, partition functions and related quantities have been computed for instantons on several spaces beyond $\BC^2$ following the seminal work of Vafa and Witten \cite{Vafa:1994tf}; examples of recent works include orbifolds $\mathbb{C}^2/\mathbb{Z}_n$ \cite{Fucito:2004ry, Bonelli:2011kv, Dey:2013fea, Ito:2013kpa} as well as general toric spaces \cite{Fucito:2006kn, Gasparim:2008ri, Bruzzo:2009uc, Bonelli:2012ny}.

In the case of $\CP^2$, the ADHM-like construction of instantons in the unitary group was developed by King in \cite{king} based on previous work by Buchdal in \cite{buchdal1,buchdal2,buchdal3} , while the constructions for special orthogonal and symplectic groups were subsequently developed by Bryan and Sanders in \cite{bryansanders2}. In this paper we review all of such constructions and their realisations from field theories which, to the best of our knowledge, have not been spelt out  anywhere in the physics literature. Interestingly, such constructions can be naturally embedded into three dimensional quiver gauge theories with 4 supercharges. For instantons of the unitary group, such a quiver theory has been introduced in \cite{Benini:2009qs} in the context of the worldvolume of M2 branes probing a Calabi-Yau 4-fold singularity (${\rm CY}_4$).  On the contrary to \cite{Benini:2009qs}, we do not focus on the full mesonic moduli space of such a theory; rather, we concentrate on a particular branch, dubbed the \textit{instanton branch}, of the moduli space on which the chiral fields acquire vacuum expectation values.  On this branch, the monopole operators do not play any role in the chiral ring.  Making use of this construction as well as the analogous ones for instantons in the orthogonal and symplectic groups, we compute the generating function, also known as Hilbert series, that counts gauge invariant quantities on the instanton branch refining them with respect to commuting $U(1)$ global charges. As a result, Hilbert series for distinct instanton configurations on $\CP^2$ turn out to be equal to those of instantons on $\BC^2$.  This equality was in fact also pointed out in \cite{Nakajima:2008eq}.
This suggests that, even though such quiver theories have only 4 supercharges and not 8 supercharges as those for the standard ADHM construction, the instanton branch possesses the expected hyperK\"ahler property of the instanton moduli space. As we point out in the main part of the paper, to see such a hyperK\"ahler structure and obtain correct Hilbert series, it is crucial to take into account large anomalous dimensions of quiver fields in such theories with 4 supercharges.  These Hilbert series are confirmed by a different method of computations, namely via the blowup formula for the Hirzebruch surface $\BF_1$ \cite{Nakajima:2005fg, Bonelli:2012ny, Keller:2012da}.

This paper is organised as follows. In section \ref{unitary} we introduce the ADHM-like construction of unitary instantons on $\CP^2$, reviewing as well some relevant aspects of the standard ADHM construction of instantons on $\BC^2$, introducing in particular several spaces which will play a role in later sections. We then embed the ADHM-like construction into a gauge theory admitting a stringy description as the worldvolume theory on M2 branes probing a certain ${\rm CY}_4$. Using the gauge theory description we compute the Hilbert series of the instanton moduli space. In section \ref{symplectic} and section \ref{orthogonal} we introduce the theories for symplectic and orthogonal instantons and use them to compute their corresponding Hilbert series. Interestingly, in all cases the Hilbert series turns out to be identical to the Hilbert series of certain instantons on $\BC^2$. This is because the set of operators relevant for the Hilbert series are in one-to-one correspondence with solutions of the corresponding ADHM construction for instantons on $\BC^2$ \cite{Nakajima:2008eq}. In section \ref{blowupformula}, we recover these results from the point of view of the blowup formula, obtaining precise agreement with our Hilbert series. We finish with some conclusions in section \ref{conclusions}.

\section{$U(N)$ instantons on $\CP^2$}\label{unitary}

We are interested in constructing $U(N)$ instantons on $\CP^2$. This has a natural relation to a problem in differential geometry as  it turns out that, on very general grounds, there is a correspondence between the moduli space of instantons on projective algebraic surfaces and the moduli space of (stable) holomorphic bundles --the so-called Hitchin-Kobayashi correspondence-- which is rather well understood in the mathematical literature \textit{e.g.} \cite{Atiyah:1979iu, DonaldsonKronheimer, king} and references therein. Hence, the problem of studying instantons reduces, in these cases, to that of constructing holomorphic  bundles. In this context, the ADHM construction can be regarded as a device to construct such bundles.

In the particular case of instantons on an $S^4$, an alternative version --which turns out to be more useful for our true purposes of studying instantons on $\CP^2$-- of the Hitchin-Kobayashi correspondence was proven by Donaldson \cite{Donaldson:1984tm, Donaldson:1985zz} by using the so-called Ward correspondence which associates an anti-self-dual (ASD) connection --that is, a connection whose curvature is ASD-- on $S^4$ to a holomorphic bundle on $\CP^2$. For completeness, and in order to introduce spaces which will be later relevant to our purposes, we will present a very rough outline here and refer to \textit{e.g.} \cite{king} for an exhaustive description. 

The key observation is that $\mathbb{C}^2$ admits two possible compactifications. On one hand, we can find a conformal compactification into $S^4$ given by the stereographic projection. Since both the Yang-Mills equations and the self-duality condition are conformally invariant, finite energy solutions on $\mathbb{C}^2$ can be extended to $S^4$ by adding the point at infinity, where the gauge field and the allowed gauge transformations become trivial.\footnote{Upon the addition of the point at infinity over which the gauge field as well as the gauge transformations are specified --in this case to be trivial-- the instanton is referred to as \textit{framed}.}  

On the other hand, we can find a holomorphic compactification into $\CP^2$ by adding the line at infinity $\ell_{\infty}$. With a little bit of hindsight, we will be interested on holomorphic bundles on $\mathbb{C}^2$. As this is a rather sick notion, the addition of $\ell_{\infty}$ allows to define them on the compact space $\CP^2$ provided we demand these bundles to be trivial, \textit{i.e.} a copy of $\mathbb{C}^{\text{rank}}$, over $\ell_{\infty}$. 

In order to construct ASD connections on $S^4$, we note that, while $S^4$ is not a complex manifold, the conformal de-compactification $\mathbb{C}^2$ is (in fact it is a K\"ahler space). Hence, considering the standard complex structure and the orientation naturally induced by it, we can can write the splitting of middle degree forms into self-dual (SD) 2-forms $\Lambda_+$ and anti-self-dual (ASD) 2-forms $\Lambda_-$ in terms of complexified forms as

\begin{equation}
\label{splitting}
\Lambda_+=\Lambda^{(2,\,0)}\oplus\Lambda^{(0,\,2)}\oplus \omega\,\Lambda^{(0,\,0)}\, ,\qquad \Lambda_-=\Lambda_0^{(1,\,1)}\, ;
\end{equation}
where $\Lambda_0^{(1,\,1)}$ stands for $(1,\,1)$ forms primitive to the K\"ahler form $\omega$, which is of course, $(1,\,1)$ SD. 

According to (\ref{splitting}), the field strength of ASD unitary connections is in $\Lambda_0^{(1,\,1)}$, the operator $\bar{\partial}_A$ (whose linearization is $\bar{\partial}_A=\bar{\partial}+\bar{A}$) satisfies that $\bar{\partial}_A^2=0$. This integrability condition ensures, through the Newlander-Nirenberg theorem, that the equation $\bar{\partial}_A\,s=0$ has the maximal number of solutions, hence defining the sections --and from there the holomorphic structure-- of a holomorphic bundle on $\mathbb{C}^2$. However, being $\mathbb{C}^2$ a non-compact space, the notion of moduli for such bundle is ill-defined. One can cure this problem by adding $\ell_{\infty}$ so that $\mathbb{C}^2$ is holomorphically compactified into $\CP^2$ while demanding the bundle to be trivial there, thus turning it into a bundle on $\CP^2$. This rough argument suggests that indeed framed ASD connections on $S^4$ are in one-to-one correspondence with holomorphic bundles on $\CP^2$ with trivial behaviour at $\ell_{\infty}$. A more precise derivation makes use of the so-called twistor space introduced by Penrose. Starting with a Riemannian manifold $M$ with metric $g$, one can associate a 2-form $\omega$ to every possible complex structure $J$ on $TM$ as $\omega=g(\cdot,\,J\,\cdot)$. This identifies the bundle of compatible complex structures with the sphere bundle in $\Lambda^2_+$. The total space $Z(M)$ of this bundle is the so-called twistor space and it turns out to be a complex manifold as long as $M$ is conformally anti-self-dual (that is, the self-dual part of the Weyl tensor vanishes, as it happens for instance on $S^4$ and $\overline{\CP}^2$, the opposite-oriented complex projective plane). Then, considering an hermitean vector bundle $\mathcal{E}$ on $M$ with a unitary connection, the Ward correspondence allows to assign to each instanton --that is, to each ASD connection of $\mathcal{E}$-- a holomorphic bundle on $Z(M)$. By appropriately considering restrictions of this bundle, Donaldson was able to prove in \cite{Donaldson:1984tm, Donaldson:1985zz} that indeed one can find a bijection between the moduli space of instantons on $S^4$ and the moduli space of holomorphic bundles on $\CP^2$ (see, for a complete explanation, \textit{e.g.} \cite{king, Atiyah:1979iu, DonaldsonKronheimer}).\footnote{To be fully precise, the relevant holomorphic bundles are demanded to be stable.} The ADHM construction can be regarded, from this point of view, as the explicit construction of such holomorphic bundles.

This analysis can be extended to the case of our interest, namely instantons on $\CP^2$. In order to build ASD connections on $\CP^2$, following \cite{buchdal1,buchdal2,buchdal3,king}  we start with $\widehat{\mathbb{C}}^2$, the blow up of $\mathbb{C}^2$ at a point defined as

\begin{equation}
\widehat{\mathbb{C}}^2=\{(x_1,\,x_2)\times [z_1,\,z_2]\,\in\,\mathbb{C}^2\times \CP^1\,/\,x_1\,z_1=x_2\,z_2\}\, .
\end{equation}
Then, by analogy with the $\mathbb{C}^2$ case, on one hand we can find a conformal compactification of $\widehat{\mathbb{C}}^2$ into $\overline{\CP}^2$ --the opposite-oriented $\CP^2$-- as follows
\begin{equation}
\label{conformalcompactification}
\widehat{\mathbb{C}}^2\,\rightarrow\,\overline{\CP}^2\,:\,\Big((x_1,\,x_2)\times [z_1,\,z_2]\Big)\,\rightarrow\,\begin{cases} [|x|^2,\,x_1,\,x_2] \, , \\ [0,\,z_1,\,z_2]\, . \end{cases}
\end{equation}
Exactly as in the $S^4$ case, we assume a trivial behavior at the added point, so that our instantons on $\overline{\CP}^2$ are framed.

On the other hand we can find a holomorphic compactification by adding the line at infinity $\ell_{\infty}$.  With hindsight, we will consider bundles over this compactification which are trivial over this line. This compactifies $\widehat{\mathbb{C}}^2$ into $\CP^2$ blown up at a point, that is, Hirzebruch's first surface $\mathbb{F}_1$. The same argument as above identifies framed ASD connections on $\overline{\CP}^2$ with holomorphic bundles on $\mathbb{F}_1$ trivial over $\ell_{\infty}$ (recall that  $\overline{\CP}^2$  is a conformally-anti-self-dual space, and hence it admits a twistor space which is a complex manifold. In this case it is the flag manifold $\mathbb{F}^{3,\,2}$). These have been constructed by King in \cite{king} using monads, who provided an ADHM-like construction for framed ASD connections on $\overline{\CP}^2$. 

Let us summarise the key ingredients of the King construction described in section 4 of \cite{king}.  The construction involves the following ingredients:
\bi
\item Three vectors spaces $V_{k_1}$, $V'_{k_2}$ and $W_N$ of dimension $k_1$, $k_2$ and $N$ respectively.
\item Homomorphisms between these spaces; denoted as follows: $B_1, B_2 \in {\rm Hom}(V'_{k_2},\,V_{k_1})$, $A_2 \in {\rm Hom}(V_{k_1},\,V'_{k_2})$, $Q \in {\rm Hom}(W_N, \, V_{k_1})$ and $q \in {\rm Hom}(V'_{k_2}, \, W_N)$.
\item The integrability equation
\bea \label{U(N)Fflat}
B_1\cdot A_2\cdot B_2-B_2\cdot A_2\cdot B_1+q \cdot Q=0~.
\eea
\ei
The solutions of \eref{U(N)Fflat} are acted by the automorphism group $G=U(k_1)\times U(k_2)$.  The King construction \cite{king} then identifies the moduli space of rank $N$ framed ASD instantons on $\overline{\CP}^2$ with the quotient of the space of such solutions by the complexification of $G$, namely $GL(k_1, \BC) \times GL(k_2, \BC)$.  As we shall explain below, these ingredients can be nicely described in terms of $3d$ $\CN=2$ quiver diagram depicted in \eref{quivSUN} and the integrability equation \eref{U(N)Fflat} being identified with the $F$-terms coming from the superpotential \eref{WSUN}.

Even though in the following we will loosely speak about instantons on $\CP^2$ \footnote{Unfortunately this plagues the literature. To begin with, the seminal contributions of \cite{buchdal1,buchdal2,buchdal3,king} refer already in the title to instantons on $\CP^2$.}, it is important to stress that the connections constructed in this way are ASD connections on $\overline{\CP}^2$. This manifold is, in a sense, not a complex manifold (since the orientation does not follow from the K\"ahler form). In fact, note that the conformal compactification in eq. (\ref{conformalcompactification}) is not a holomorphic map, since it involves $|x|^2$. Of course, upon inversion of the orientation we obtain SD connections on the standard-oriented $\CP^2$, where the K\"ahler form is SD. Hence the King construction can be regarded as that of instantons on $\CP^2$ with the same duality properties as the K\"ahler form. In particular, this implies that our instantons, regardless on other issues such as framing,  are just the opposite to the ones appearing in \cite{Kim:2012qf}. In fact, as opposed to the case in \cite{Kim:2012qf}, where the instanton partition functions are computed by ``patching" the contributions from three fixed points, as we will show in section \ref{blowupformula}, \textit{via} the blowup formula \eref{blowup}, our instanton partition functions receive contributions from only two ``patches''\footnote{These are labelled by $r=0$ and $r=1$ in \eref{blowup}.}, each of which contains the partition function of instantons on $\mathbb{C}^2$. The physical significance of this remains yet to be fully exploited.

In the following we will embed the King ADHM-like construction in a gauge theory engineered by branes probing a certain singularity and use it to compute the Hilbert series of the instanton moduli space. We refer to \cite{king} for further details on the ADHM-like construction of instantons on $\CP^2$.

\subsection{Gauge theory for the King construction}
The construction for $U(N)$ instantons on $\mathbb{CP}^2$ proposed by \cite{king} can be realised from a $3d$ $\CN=2$ gauge theory whose quiver diagram is depicted in \eref{quivSUN} with the superpotential
\bea \label{WSUN}
W&= (A_1)^{a_1}_{~a_2} (B_1)^{a_2}_{~b_1} (A_2)^{b_1}_{~b_2} (B_2)^{b_2}_{~a_1} - (A_1)^{a_1}_{~a_2} (B_2)^{a_2}_{~b_1} (A_2)^{b_1}_{~b_2} (B_1)^{b_2}_{~a_1} + Q^{i}_{~a_1} (A_1)^{a_1}_{~a_2} q^{a_2}_{~i}   \nn \\
&=  \epsilon^{\alpha \beta}\,\epsilon^{\sigma \rho} \; \mathrm{Tr} (A_ \alpha \cdot B_\sigma \cdot A_\beta \cdot B_\rho )+Q^i \cdot A_1\cdot q_i~,
\eea
where $a_1, b_1, \ldots =1, \ldots, k_1$ are the gauge indices for $U(k_1)$; $a_2, b_2, \ldots =1, \ldots, k_2$ are the gauge indices for $U(k_2)$; $i,j =1, \ldots, N$ are the flavour indices for $U(N)$.
\bea 
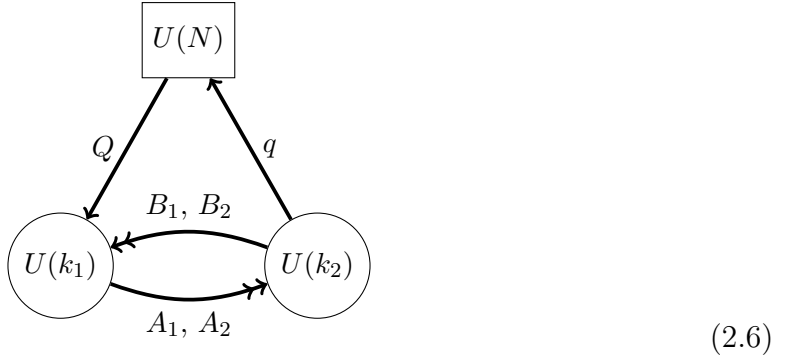
\label{quivSUN}
\begin{tikzpicture}[font=\small]
\begin{scope}[auto,%
  every node/.style={draw, minimum size=1cm}, node distance=2cm];
\node[circle] (k1) at (0,0) {$U(k_1)$};
\node[circle, right=of k1] (k2) {$U(k_2)$};
\node[rectangle] (N) at (1.7,3) {$U(N)$};
\end{scope}
\draw[draw=black,solid,line width=0.5mm,->>] (k1) to[bend right=20] node[midway,above] {}
      node[midway,below] {$A_1$, $A_2$} (k2);
\draw[draw=black,solid,line width=0.5mm,<<-] (k1) to[bend left=20] node[midway,above] {$B_1$, $B_2$}
      node[midway,below] {} (k2);
\draw[draw=black,solid,line width=0.5mm,->] (k2) to node[midway,right] {$q$}  (N);    
\draw[draw=black,solid,line width=0.5mm,->] (N) to node[midway,left] {$Q$}  (k1);    
\end{tikzpicture}
\eea
Without the flavors the quiver exhibits an $SU(2)\times SU(2)$ global symmetry, indicated by the indices $\{ \alpha, \beta=1,2\}$ and $\{\sigma, \rho=1,2\}$ respectively. After coupling $A_1$ to the $U(N)$ flavour symmetry, such a symmetry is broken to the second factor of $SU(2)$ that rotates the $B_\sigma$ fields.

As we will discuss in more detail below, the King construction arises in the branch of the moduli space where we set $A_1=0$. Then, the only non-trivial F-term is precisely 
\bea \label{FtermsSUN}
0=\frac{\partial W}{\partial {A_1}^{a_1}_{~a_2}} =   \left( B_1 \cdot A_2 \cdot B_2 - B_2 \cdot A_2 \cdot B_1 \right)^{a_2}_{~a_1} +  q^{a_2}\cdot Q_{~a_1}  ~.
\eea
This is indeed the integrability equation given by \eref{U(N)Fflat} and Eq. (I) of section 4 in \cite{king}. 
Upon quotienting by the complexification of the gauge symmetry, we precisely find the moduli space of $U(N)$ instantons on $\CP^2$.

In addition to the ingredients stated above, King's construction involves further constraints coming from the stability conditions. In physics language these map to restricting the FI parameters of the ADHM-like quiver in the  to live in certain domains. We refer to \cite{king, Nakajima:2008eq} for further details. Nevertheless, for our purposes, since we will be interested on computing the Hilbert series of the corresponding instanton moduli spaces, these signs play no role.

Note that the quiver diagram \eref{quivSUN}, interpreted as a four dimensional gauge theory, would contain a gauge anomaly. Nevertheless we can still make sense of this gauge theory in three dimensions at the expense of non-zero bare Chern-Simons (CS) levels. Indeed, gauge invariance requires the CS levels to be\footnote{The condition on the bare CS level $\kappa$ is $\kappa+ \frac{1}{2} \sum_f d_3(R_f) \in \BZ$, where the sum is over fermion $f$ in the representations $R_f$, and $d_3(R)$ is the cubic index of $R$.  Here $d_3$ for the fundamental and antifundamental representations of $SU(k)$ are $+1$ and $-1$ respectively.} $N/2+\kappa_1$ and $-N/2+\kappa_2$ respectively, with $\kappa_1$ and $\kappa_2$ integers. Indeed, this theory is among the families considered in \cite{Benini:2009qs}, namely flavoured ABJM theories. It is important to stress that these theories are strongly coupled conformal field theories.


An interesting feature of the King construction is the fact the superpotential $W$ is quartic and hence the $F$-terms \eref{FtermsSUN} are cubic. This strongly suggests that the corresponding gauge theory will have 4 supercharges. This is in sharp contrast with the embedding of the ADHM construction for instantons on $\mathbb{C}^2$ into a gauge theory with 8 supercharges. In particular, that amount of SUSY does not allow large anomalous dimensions for the fields, whose dimension is then just the classical one. At the same time, the moduli space splits into a Coulomb and a Higgs branch, the latter encoding the instanton moduli space and free of quantum corrections. In contrast, in the case at hand, owing to the low SUSY, the fields will develop large anomalous dimensions and  the dynamics of the gauge theory and the structure of its moduli space will be in general very complicated. Suppose for example embedding less minimally the King construction into a 4d $\mathcal{N}=1$ gauge theory, considering for instance the flavoured conifold as described in \cite{Benini:2007kg}. Upon setting $A_1=0$, $q=0$ and $\tilde{Q}=0$ in (2.6) in that reference we would have obtained the same $F$-term relations and gauge constraints. Nevertheless, this $\mathcal{N}=1$ theory is not conformal and, depending on the ranks of flavour and colour nodes, the IR dynamics can be very different, in particular breaking any $R$-symmetry.\footnote{Note however that we could imagine a regime in which the two colour ranks are the same while the flavour rank is much smaller, so that we can assume being sufficiently close to the Klebanov-Witten fixed point. Under that assumption, we can take the dimensions of the bifundamental fields to be those of the KW theory and fix the quark dimensions by $a$-maximisation. Even though we will not pursue this further in the following, one can see that this alternative dimension assignation would lead to the same quantitative results as below --\textit{i.e.} the same Hilbert series-- upon appropriate rescalings of fugacities. However, this is trustable only in a very special rank and energy regime, and hence we prefer to concentrate on the 3d embedding into a conformal theory from now on.} In contrast, the proposed 3d theory is a CFT and hence has a controlled behavior. Because of this, it is more natural to embed the construction of $\CP^2$ instantons in a three dimensional theory, which is conformal and admits a preferred superconformal $R$-symmetry grading the instanton branch. 

Since the gauge theory we will consider is three dimensional, there are monopole operators that can affect the chiral ring of the theory.  For example, as discussed in (6.15) of \cite{Benini:2009qs}, in the case of $k_1=k_2=1$ and $\kappa_1=-\kappa_2=\kappa$, there are monopole operators $T$ and $\widetilde{T}$ that carry gauge charges $(\kappa+N, -\kappa-N)$ and $(-\kappa, \kappa)$ under $U(k_1)\times U(k_2)$ respectively; moreover, $T$ and $\widetilde{T}$ satisfy a quantum relation $T \widetilde{T}=A_1^N$ and both $T$ and $\widetilde{T}$ carry $U(1)$ $R$-charges $N/2$.  These pieces of information are crucial for the computation of the $R$-charges of each chiral field. However, as we shall discuss subsequently, the monopole operators do not play a role in our analysis on the moduli space of instantons on $\CP^2$.

\subsubsection*{The charge assignments}

As discussed above, it is natural to embed the King construction into a 3d strongly coupled CFT with 4 supercharges. Hence we need to determine the preferred choice of superconformal R-symmetry. To that matter, we summarise the transformations of each chiral field and monopole operator under the gauge and global symmetries in \tref{tab:chargesSUN}.  Note that the R-charges are fixed by the following requirements: 
\ben
\item  The superpotential \eref{WSUN} carries an $R$-charge 2.
\item  The special case of $k_1=k_2=1$, $N=1$ and $\kappa_1= -\kappa_2 =1/2$. As shown in section 6.2.1 of \cite{Benini:2009qs}, the mesonic moduli space is a conifold times a complex line.  The conifold is parametrized by four fields $x_1=T B_1$, $x_2=A_2 B_2$, $x_3=T B_2$, $x_4 = A_2 B_1$, subject to the relation $x_1 x_2-x_3 x_4=0$; the complex line is parametrised by the monopole operator $\widetilde{T}$.
\ben
\item According to the charge assignments and symmetry argument given in section 4 of \cite{Davey:2009sr}, each field $x_1, \ldots x_4$ that parametrizes the conifold carries the same $R$-charge $3/4$ and the field $\widetilde{T}$ that parametrizes the complex line, as a free field, carries the $R$-charge $1/2$.
\item Imposing the conditions that $T$ and $\widetilde{T}$ carry equal $R$-charges $1/2$, that $B_1, \; B_2$ carry equal $R$-charges (due to their transformation as a doublet of an $SU(2)$ global symmetry) and that $T \widetilde{T}=A_1$, we obtain the $R$-charges of $A_1, A_2, B_1, B_2$ as in \tref{tab:chargesSUN}. 
\item We conjecture that these $R$-charge assignments hold for any rank of the gauge groups $k_1$ and $k_2$ and the flavour symmetry $U(N)$.
\een
\item The $R$-charges of $Q$ and $q$ are left undetermined.  The consistency with the above assignments requires them to have charges $\frac{1}{4}r$ and $1-\frac{1}{4}r$ respectively\footnote{The normalisation $1/4$ is introduced for convenience.}.  We shall see that the unknown value of $r$ does not affect the subsequent analysis of the moduli space of instantons.
\een

\begin{table}[H]
\begin{center}
\begin{tabular}{|c||c|c||c|c|c|}
\hline
~ & $U(k_1)$ & $U(k_2)$ & $SU(2)$ & $U(N)$ & $U(1)_R$ \\
\hline
$A_1$ & $[1,0,\ldots,0]_{+1}$ &$[0,\ldots,0,1]_{-1}$ &$[0]$ & $[\vec 0]_0$& 1 \\
$A_2$ & $[1,0,\ldots,0]_{+1}$ &$[0,\ldots,0,1]_{-1}$& [0] & $[\vec 0]_0$ & $1/2$ \\
$B_1,\; B_2$ & $[0,\ldots,0,1]_{-1}$ &$[1,0,\ldots,0]_{+1}$ & [1] & $[\vec 0]_0$& $1/4$ \\
$Q$ & $[0,\ldots,0,1]_{-1}$ & $[\vec 0]_0$ & [0] & $[1,0,\ldots,0]_{+1}$& $\frac{1}{4}r$ \\
$q$ &$[\vec 0]_0$ & $[1,0,\ldots,0]_{+1}$ & [0] & $[0,\ldots,0,1]_{-1}$ &  $1-\frac{1}{4}r$ \\
\hline

\end{tabular}
\end{center}
\caption{Charge assignments of the chiral fields in theory \eref{quivSUN}.}
\label{tab:chargesSUN}
\end{table}%

\subsubsection*{The instanton branch of the moduli space}

As mentioned above, roughly speaking the King construction of instantons  on $\CP^2$ arises as the $A_1=0$ branch of the moduli space. However, due to the presence of monopole operators, we need to be more precise. In fact, according to the construction described in \cite{king}, the branch of the moduli space that can be identified with the instanton moduli space is the one on which $A_1=T=\widetilde{T}=0$, which is consistent with the quantum relation among monopoles on the chiral ring.  We shall henceforth refer to this as an {\it instanton branch}.  On this branch, the relevant $F$-terms are given by (\ref{FtermsSUN}).

Note that the description of this branch of the moduli space shares some similarity to that of the Higgs branch of the standard ADHM construction.  It is parametrized by the the gauge invariant quantities constructed from the massless chiral fields $\{ A_2, B_1, B_2, Q, q\}$ subject to constraints from the $F$-terms.  We shall henceforth denote the instanton branch of quiver \eref{quivSUN} by ${\mathcal{M}}^{SU(N)}_{\CP^2}$ and subsequently show that this space is a cone endowed with a hyperK\"ahler structure.  It should be emphasised that this branch of moduli space is different from the mesonic moduli space described in \cite{Benini:2007kg}.

\subsection{The Hilbert series of the instanton branch}
The Hilbert series of $U(N)$ instantons on $\CP^2$ with the configuration $\vec k =(k_1, k_2)$ can be obtained by first computing the Hilbert series of the space of the $F$-term solutions, also known as the $F$-flat ($\fflat$) space\footnote{It can be checked case by case using an algebraic software, \eg~ Macaulay2 \cite{M2}, that the $F$-flat space associated with \eref{FtermsSUN} is a complete intersection. As a consequence, there are a finite number of terms in the $\PE$ in the integrand of \eref{HS}; the terms with positive contributions correspond to the generators and that with the negative contribution corresponds to the relation. \label{CIremark}}, and then integrating it over the Haar measure of the gauge symmetry.  Explicitly, this is given by
\bea
\label{HS}
H[\vec k, U(N), \CP^2](t; x; \vec y) &= \int \ud \mu_{U(k_1)} (\vec u)  \int \ud \mu_{U(k_2)} (\vec w) \times\nn \\
& \PE \Big[ \chi_{A_2} t^2+ \chi_{B_1,B_2} t+ \chi_{Q} t^r +\chi_q t^{4-r} -  \chi_{\text{F-terms}} t^4\Big]~,
\eea
where the notations are explained below:
\bi
\item The fugacity $t$ keeps track of the $R$-charge in the unit of $1/4$.
\item The fugacities $\vec u$ and $\vec w$ are associated with the gauge groups $U(k_1)$ and $U(k_2)$, respectively.
\item The fugacities $x$ and $\vec y$ are associated with the global symmetries $SU(2)$ and $SU(N)$, respectively.
\item The contributions from each chiral field are given below:
\bea
\begin{array}{ll}
 \chi_{A_2} = \sum_{a=1}^{k_1} \sum_{a=1}^{k} u_a w_b^{-1}  ~,  & ~\chi_{B_1,B_2} = ( x+x^{-1}) \sum_{a=1}^{k_1} \sum_{b=1}^{k_2} u_a^{-1}  w_b ~,   \\
 \chi_{Q} =\left(\sum_{a=1}^{k_1} u^{-1}_a \right) \left( \sum_{i=1}^N y_i\right)~, & ~\chi_{q} = \left( \sum_{a=1}^k w_a \right) \left(  \sum_{i=1}^N y_i^{-1} \right)~, \\
 \chi_{\text{F-terms}} = \sum_{a=1}^{k_1} \sum_{b=1}^{k_2} u^{-1}_a  w_b~.
\end{array}
\eea
\item The Haar measure of $U(k)$ can be taken as
\bea
\int \ud \mu_{U(k)} (\vec w) = \frac{1}{N!} \left( \prod_{j=1}^k \oint_{|w_j|=1} \frac{\ud w_j}{2 \pi i w_j} \right) \prod_{1\leq i <j \leq k} (w_i-w_j)(w_i^{-1}-w_j^{-1})~.
\eea
\ei

We demonstrate in a number of examples below that the Hilbert series of $U(N)$ instantons on $\CP^2$ with the configuration $\vec k=(k_1, k_2)$ is equal to the Hilbert series of $SU(N)$ instantons on $\BC^2$ with instanton number $\min(k_1,k_2)$:
\bea
H[(k_1,k_2), U(N), \CP^2](t; x; \vec y) = H[\min(k_1,k_2), SU(N), \BC^2] (t^3; x; \vec y)~, \label{mink1k2}
\eea
where the Hilbert series $H[k, SU(N), \BC^2](t; x; \vec y)$ of $k$ $SU(N)$ instantons on $\BC^2$ is studied in \cite{Benvenuti:2010pq, Hanany:2012dm}.  This relation was in fact proven in \cite{Nakajima:2008eq}.

Note that the result is symmetric under the exchange of $k_1$ and $k_2$. This stands for the fact that the quiver is invariant under the combined action of exchanging the gauge groups and charge conjugation.

\subsection*{Mapping between the King construction and the ADHM construction}

Following section 1.3 in  \cite{Nakajima:2008eq}, it is possible to understand \eref{mink1k2} by considering a map between the King construction and the ADHM construction. Let us assume for definiteness that $k_1\leq k_2$ and define the map (see section 1.3 in  \cite{Nakajima:2008eq})
\begin{equation} \label{mappi}
\pi:\, (A_2,\,B_1, \, B_2, \,Q,\,q)\rightarrow (\widehat{X}_1 = A_2\, B_1,\,\widehat{X}_2 = A_2\,B_2,\,\widehat{I} =A_2\, q ,\,\widehat{J} =Q)
\end{equation}
where $\widehat{X}_{1,2}$ transform under the adjoint representation of $k_1 = \min (k_1,k_2)$, $\widehat{I}$ is a bifundamental chiral field of $U(k_1) \times U(N)$, and $\widehat{J}$ is a bifundamental chiral field in $U(N) \times U(k_1)$.  After being multiplied by $A_2$ on the left, the $F$-terms \eref{FtermsSUN} is mapped by $\pi$ into the following relation:
\bea \label{ADHMFterms}
[\widehat{X}_1,\,\widehat{X}_2]+\widehat{I}\cdot \widehat{J}=0\, .
\eea
Indeed, $\pi$ maps the King construction for $U(N)$ instantons on $\CP^2$ into the ADHM construction for $U(N)$ instantons on $\BC^2$. Here $\widehat{X}_1, \, \widehat{X}_2, \,\widehat{I},  \,\widehat{J}$ are chiral fields appearing in the ADHM quiver \eref{ADHMquiv} for $k$ $U(N)$ instantons on $\BC^2$, written in terms of $4d$ $\CN=1$ notation, where $k= \min(k_1, k_2)$.
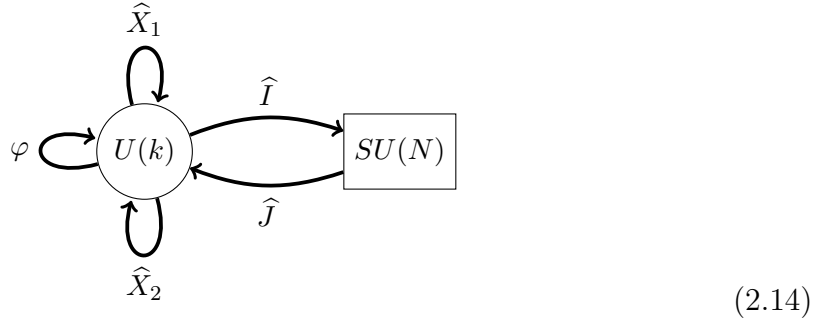
\bea \label{ADHMquiv}
\begin{tikzpicture}[font=\small]
\begin{scope}[auto,%
  every node/.style={draw, minimum size=1cm}, node distance=2cm];
\node[circle] (k) at (0,0) {$U(k)$};
\node[rectangle, right=of k] (N) {$SU(N)$};
\end{scope}
\draw[draw=black,solid,line width=0.5mm,->] (k) to[bend left=20] node[midway,above] {$\widehat{I}$}
      node[midway,below] {} (N);
\draw[draw=black,solid,line width=0.5mm,<-] (k) to[bend right=20] node[midway,above] {}
      node[midway,below] {$\widehat{J}$} (N);
\draw[draw=black,solid,line width=0.5mm,->]    (k) edge [loop above] node {$\widehat{X}_1$} (k) ;
\draw[draw=black,solid,line width=0.5mm,->]    (k) edge [loop left] node {$\varphi$} (k) ;
\draw[draw=black,solid,line width=0.5mm,->]    (k) edge [loop below] node {$\widehat{X}_2$} (k) ;
\end{tikzpicture}
\eea 
The quiver diagram depicted in \eref{ADHMquiv} has the superpotential
\bea
W_{\text{ADHM}} = \tr\left( \widehat{J} \cdot \varphi \cdot  \widehat{I} + \varphi \cdot [\widehat{X}_1, \, \widehat{X}_2] \right)~.
\eea
The F-terms relevant to the Higgs branch of the ADHM quiver \eref{ADHMquiv} are given by $\partial_{\varphi} W_{\text{ADHM}}=0$, which is precisely \eref{ADHMFterms}. 

On the other hand, it is also possible to construct a map that embeds the ADHM construction into the King construction, namely (see section 1.3 in  \cite{Nakajima:2008eq})
\bea
\sigma: \;  (\widehat{X}_1,\,\widehat{X}_2,\,\widehat{I}  ,\,\widehat{J} ) ~ \rightarrow ~ (A_2= {\bf 1}_{k \times k}, \, B_1 =\widehat{X}_1,\,B_2 = \widehat{X}_2,\,q = \widehat{I}   ,\,Q = \widehat{J} )~, 
\eea
where ${\bf 1}_{k \times k}$ denotes the $k \times k$ identity matrix. This is accompanied with the obvious extension of the gauge group. It is clear that the $F$-terms \eref{ADHMFterms} is mapped by $\sigma$ into \eref{FtermsSUN}.  Indeed, $\sigma$ induces the hyperK\"ahler structure from the moduli space of instantons on $\BC^2$ to the moduli space of those on $\CP^2$.


As a consequence of these maps, the algebraic constructions of the moduli space of $\vec k = (k_1, k_2)$ and the moduli space of $\min (k_1, k_2)$ $U(N)$ instantons on $\BC^2$ are identified. This algebraic identification stands behind the Hilbert series identity found above.  Moreover

\bea \label{dimUNinstCP2}
\dim_\BC\,\mathcal{M}^{U(N)}_{\CP^2}= 2kN , \qquad k={\rm min}(k_1,\,k_2)~,
\eea

Before ending this subsection, we comment on the following crucial points.
\ben
\item As a result of the map $\pi$, the hyperK\"ahler structure for the moduli space of instantons on $\CP^2$ is established. Given that the gauge theory \eref{quivSUN} has only 4 supercharges and not 8 supercharges as for the ADHM quiver, it is not {\it a priori} clear that the expected properties of the moduli space would emerge. Nevertheless, the instanton branch ${\mathcal{M}}^{SU(N)}_{\CP^2}$ recovers the hyperK\"ahler structure through the maps we have described. 
\item In order to obtain the Hilbert series that precisely matches that of instantons on $\mathbb{C}^2$, it is crucial to use a suitable grading associated with the fugacity $t$ that is compatible with the correct superconformal R-symmetry. We present these $R$-charges in \tref{tab:chargesSUN}.
\een

\subsection{Properties of the instanton moduli space}

The moduli space of $U(N)$ instantons on $\mathbb{C}^2$ is characterised by the rank $N$ of the gauge bundle as well as by its instanton number. It is natural to search for the analogous characterization of the moduli space of instantons of the unitary group on  $\CP^2$. 

A priori, unitary instantons on $\CP^2$ are characterized by the rank of the gauge group as well as by their Chern classes. Since $\CP^2$ is a topologically non-trivial space, in contrast to the $\mathbb{C}^2$ case, we should expect both the first and second Chern classes to be non-trivial. Hence, the instanton moduli spaces of interest are characterized by three numbers, which must then correspond to certain combinations of the quiver data $k_1$, $k_2$, and $N$. 

One option is to characterise the instanton bundle by the first Chern number $\hat{c}$ and the instanton number $\hat{k}$.\footnote{We shall henceforth reserve the hatted symbols for the instanton data and unhatted symbols for the quiver data.}  In terms of the topological data of the corresponding holomorphic bundle $E$ on $\mathbb{F}_1$, these are (see \textit{e.g.} \cite{Nakajima:2008eq})
\begin{equation} \label{instdata}
\langle c_1(E),\,[C]\rangle=-\hat{c}\, ,\qquad \langle c_2(E)-\frac{N-1}{2\,N}\,c_1(E)^2,\,[\mathbb{F}_1]\rangle=\hat{k}\, ;
\end{equation}
where $[C]$ is the class of $\CP^1$ inside $\mathbb{F}_1$. Besides, further explanation leading to the second definition can be found in  \cite{Witten:2000nv}.

These numbers are related to the quiver data $k_1$, $k_2$ and $N$ appearing in the King construction as follows \cite{Nakajima:2008eq}:
\begin{equation} \label{chatkhat}
\hat{c}=k_2-k_1~,\qquad \hat{k}=\frac{1}{2}(k_1+k_2)-\frac{1}{2N}(k_2-k_1)^2\, .
\end{equation}
In principle $\hat{c}\,\in\,\mathbb{Z}$ and, without loss of generality, we assume $\hat{c}\geq 0$. Note that $\hat{k}\geq 0$ needs {\it not} be an integer.  
Note that the particular case of $k_1=k_2$, \ie~the zero first Chern number, was considered in \cite{king, bryansanders1, bryansanders2}.

The relation between $k=\min(k_1,k_2)$, appearing in \eref{dimUNinstCP2}, and $\hat{k}$ can be computed by solving simultaneously the two equations in \eref{chatkhat}:
\begin{equation}
k= \min(k_1, k_2) =\hat{k}-\frac{1}{2N} \hat{c}(N-\hat{c})\, . \label{kkhat}
\end{equation}
 
\subsubsection*{The resolved moduli space of instantons}

As we have seen above, the Hilbert series counts the gauge invariant quantities on the instanton branch of the King quiver, and, thanks to the maps $\pi$ and $\sigma$, this branch coincides, as an algebraic variety, with the Hilbert series of $SU(N)$ instantons on $\mathbb{C}^2$ with an instanton number $k={\rm min}(k_1,\,k_2)$.   


The above algebraic identification does not mean that the two moduli spaces --that of instantons on $\mathbb{C}^2$ and that of instantons on $\mathbb{C}P^2$-- are equal as Riemann surfaces. Moreover, already at the algebraic level they could be different by considering more refined objects such as baryon-like branches (whose existence in general relies on appropriate choices of FI parameters). Hence, including these directions, one can think of a \textit{resolved moduli space}\footnote{By ``resolved'', we mean that it discerns these extra directions.} denoted by $\widehat{\mathcal{M}}^{U(N)}_{\CP^2}$; see Section 3 of \cite{Nakajima:2003pg}, Section 2 of \cite{Nakajima:2005fg} and Definition 1.3 of \cite{Nakajima:2008eq}.  The resolved moduli space is a non-singular space whose complex dimension can be computed by considering $3k_1k_2+k_1N+k_2N$ degrees of freedom of the chiral fields $\{A_2,\,B_1, B_2,\,Q,\,q\}$ restricted by $k_1 k_2$ $F$-term conditions and $k_1^2+k_2^2$ $D$-term conditions:

\bea \label{diffdim}
{\rm dim}_{\mathbb{C}}\,\widehat{\mathcal{M}}^{U(N)}_{\CP^2} &=3k_1k_2 +k_1 N +k_2 N -k_1 k_2 -(k_1^2+k_2^2) \nn \\
&= (k_1+k_2)N-(k_1-k_2)^2 \nn\\
&= 2\hat{k}N\, ,
\eea
in accordance with Section 2.1 of \cite{Nakajima:2005fg}. The dimension of the resolved moduli space is related to that of the instanton branch by
\begin{equation}
 {\rm dim}_{\mathbb{C}}\,\widehat{\mathcal{M}}^{U(N)}_{\CP^2}={\rm dim}_{\mathbb{C}}\,\mathcal{M}^{U(N)}_{\CP^2}+\hat{c}\,(N-\hat{c})\ .
\end{equation}
This suggests that the extra directions are associated to $\hat{c}$, which is related to dissolved surface operators associated with $\hat{c}\sim \int_{{\CP}^1} F$.  We expect that the moduli space of such operators should account for the extra directions. We will leave the study of this connection for future work and provide certain comments on this point in section \ref{conclusions}.

Note that for $\hat{c}=0$ and $\hat{c}=N$, the dimensions of both spaces are equal. 
This suggests that $\hat{c}$ should be thought as an integer modulo $N$. To understand this, let us go back to the case of $k$ $U(N)$ instantons on $\mathbb{C}^2$, whose ADHM construction \eref{ADHMquiv} consists on a $U(k)$ gauge group and a $U(N)$ flavour symmetry. However, the $U(1)$ subgroup of $U(N)$ is really gauged, as the former can be identified with the $U(1)$ subgroup of $U(k)$.  Hence the flavour symmetry of the theory is $U(N)/U(1)$.  Since $U(N) = (U(1) \times SU(N))/\BZ_N$, the group for the instanton bundle is really $SU(N)/\BZ_N$.  The same observation holds in the case at hand. Then, the $N$-ality of the instanton can be then thought as the reduction mod $N$ of the first Chern class, which naturally leads to the fact that $\hat{c}$ is really a modulo-$N$ quantity.    We shall henceforth take $\hat{c}$ to be an integer such that $0 \leq \hat{c} \leq N-1$.

Note that (\ref{diffdim}) indeed naturally suggests to interpret $\hat{k}$ as an instanton number for $SU(N)$ instantons on $\CP^2$.  On the other hand, the pararameter $k = \min (k_1, k_2)$ obtained from the King quiver \eref{quivSUN} should be viewed as the instanton number for the bundle of $SU(N)$ instantons on $\BC^2$ which is the image of the map $\pi$ given by \eref{mappi}.

\subsection{Explicit examples}

Let us now check in explicit examples our previous claim, namely that the Hilbert series for unitary instantons on $\CP^2$ coincides with the Hilbert series of ${\rm min}(k_1,\,k_2)$ $U(N)$ instantons on $\mathbb{C}^2$. 

\subsubsection{The configurations with a vanishing first Chern number: $k_1=k_2$}

As a warm-up exercise, let us first consider the configurations in which $k_1=k_2=k$ (the case of $k_1=k_2$ was introduced in \cite{king, bryansanders1, bryansanders2}).  This corresponds to the instanton bundle with the vanishing first Chern number, that is, $\hat{c}=0$. Hence in this case $\hat{k}=k$.

\subsubsection*{One $U(1)$ instanton: $\vec k =(1,\,1)$, and $N=1$}
Eq. (\ref{HS}) reads
\bea
& H[\vec k =(1,\,1),\,U(1),\,\CP^2](t, x) \nn \\
&= 
\oint_{|u|=1} \frac{\ud u}{2 \pi i u} \oint_{|w|=1} \frac{\ud w}{2 \pi i w} \PE \left[ t^2\,\frac{u}{w}+t\,(x+x^{-1})\,\frac{w}{u}+t^2\,\left(\frac{y}{u}+\frac{w}{y} \right)-\frac{w}{u}\,t^4 \right]~,
\eea
where we have chosen for definitness $r=2$. Integrating over the $U(1)\times U(1)$ gauge symmetry we find
\begin{equation}
H[\vec k =(1,\,1),\,U(1),\,\CP^2](t,x)=\frac{1}{(1-t^3\,x)\,(1-t^3\,x^{-1})}
\end{equation}
which coincides with the Hilbert series of $\mathbb{C}^2$, which is the moduli space of $1$ $U(1)$ instanton on $\BC^2$.

We can easily understand this counting directly constructing the operators. In this case, the F-terms \eref{FtermsSUN} becomes $Q\,q=0$, which demands either $Q$ or $q$ to vanish.  We therefore have two branches of the moduli space: One on which $Q=0$ and the other on which $q=0$. The branch on which $Q=0$ is freely generated by
\bea
\label{ops}
M_1=  A_2 B_1~, \quad M_2 = A_2 B_2~,
\eea
in the doublet of the $SU(2)$ global symmetry.  Similarly for the $q=0$ branch.

\subsubsection*{One $U(2)$ instanton, $\vec k =(1,1)$, and $N=2$}
Using (\ref{HS}), we find that
\begin{equation}
H[(1,1),\,U(2),\,\CP^2](t, x, \vec y) =\frac{1}{(1-t^3\,x)\,(1-t^3\,x^{-1})} \times \frac{(1+t^6)}{(1-t^6\,\frac{y_1}{y_2})\,(1-t^6\,\frac{y_2}{y_1})}\, .
\end{equation}
This is indeed the Hilbert series of $\BC^2 \times \BC^2/\BZ_2$, which is the moduli space of 1 $SU(2)$ instanton on $\BC^2$.

\subsubsection*{Two $U(1)$ instanton, $\vec k =(2,\,2)$, and $N=1$}
Using (\ref{HS}), we find that
\bea
&H[(2,\,2),\,U(1),\,\CP^2](t, x, \vec y)  \nn\\
&=\frac{(1+t^6)}{(1-t^3\,x)\,(1-t^3\,x^{-1})\,(1-t^6\,x^2)\,(1-t^6\,x^{-2})}\,   \\
&= \frac{1}{2} \left \{ H[(1,\,1),\,U(1),\,\CP^2](t, x, \vec y)^2 + H[(1,\,1),\,U(1),\,\CP^2](t^2, x^2, \vec y^2) \right \}~. \nn
\eea
This is indeed the second symmetric power of the Hilbert series of $\BC^2$, which is the moduli space of $1$ $U(1)$ instanton on $\BC^2$.

\subsubsection*{Two $U(2)$ instanton, $\vec k =(2,\,2)$, and $N=2$}
Unrefining for simplicity, we now find
\bea
& H[(2,\,2),\,U(2),\,\CP^2](t, \vec x = \vec 1, \vec y = \vec 1) \nn \\
&=\frac{1+t^3+3\,t^6+6\,t^9+8\,t^{12}+6\,t^{15}+8\,t^{18}+6\,t^{21}+3\,t^{24}+t^{27}+t^{30}}{(1-t^3)^2\,(1-t^3)^6\,(1+t^3)^4\,(1+t^3+t^6)^3}\, ,
\eea
which is equal to the Hilbert series for 2 instantons on $SU(2)$ on $\mathbb{C}^2$ given in (3.16) of \cite{Hanany:2012dm}, upon the rescaling $t \rightarrow t^3$.

\subsubsection{The configurations with a non-vanishing first Chern class: $k_1 \neq k_2$}

We now turn to the case of $k_1\ne k_2$, in which the first Chern class of the instanton bundle is non-zero.

\subsubsection*{One $U(N)$ instanton: $\vec k= (1,k)$ or $(k,1)$ with $k\geq 1$}

The result is the Hilbert series of one $SU(N)$ instantons on $\CP^2$ with the configuration $\vec k =(1,k)$:
\bea \label{HS1kSUN}
H[(1, k); SU(N)] (t; x; \vec y) &= \oint_{|u|=1} \frac{\ud u}{2 \pi i u} \int \ud \mu_{U(k)} (\vec w) H[\fflat](t; u ; \vec w ;x; \vec y)  \nn \\
&= \PE[ (x+x^{-1}) t^3 ] \left( \sum_{m=1}^\infty [m,0,\ldots,0,m]_{\vec y} t^{6m} \right)~.
\eea
This is indeed the Hilbert series of one $SU(N)$ instantons on $\BC^2$; see (3.11) of \cite{Benvenuti:2010pq}.

\subsubsection*{The generators of the moduli space}

The first factor $\PE[ (x+x^{-1}) t^3 ]$ is the Hilbert series of $\BC^2$.  The space $\BC^2$ is parametrised by the gauge invariant quatities:
\bea
(A_2)_{a_2} (B_1)^{a_2}~, \qquad (A_2)_{a_2} (B_2)^{a_2}~,
\eea
where here and throughout this section we drop the $U(1)$ gauge indices $a_1, b_1 =1$.

The second factor in \eref{HS1kSUN}, which is an infinite sum, corresponds to the reduced instanton moduli space.  The generators of this space are
\bea
M^{i}_{~j} = Q^i_{~a_1} (A_2)^{a_1}_{~a_2} q^{a_2}_{~j}~, 
\eea
with~$i,j =1, \ldots, N$. In order to see that $M$ is traceless, namely
\bea \label{trMeq0}
M^i_{~i} =0~,
\eea
we multiply $(A_2)^{a_1}_{~a_2}$ to \eref{FtermsSUN} and use the cyclic property of the trace. Indeed, $M$ transform under the adjoint representation of $SU(N)$.

Moreover, let us show that $M$ is a nilpotent matrix of degree 2, namely $M^2=0$. Consider the elements of the matrix $M^2$:
\bea \label{M2eq0}
(M^2)^{i}_{~k} &= M^{i}_{~j} M^{j}_{~k}  \nn \\
&=  Q^i(A_2)_{a_2} q^{a_2}_{~j}Q^j (A_2)_{a'_2} q^{a'_2}_{~k} \nn \\
&\stackrel{F}{=}  -Q^i (A_2)_{a_2} \left( B_1 \cdot A_2 \cdot B_2 - B_2 \cdot A_2 \cdot B_1 \right)^{a_2} (A_2)_{a'_2} q^{a'_2}_{~k} \nn \\
&= - Q^i \left( A_2 \cdot B_1 \cdot A_2 \cdot B_2  - A_2 \cdot B_2 \cdot A_2 \cdot B_1 \right) (A_2)_{a'_2}  q^{a'_2}_{~k} \nn \\
&= -Q^i [A_2 \cdot B_1,\; A_2 \cdot B_2] (A_2)_{a'_2} q^{a_2'}_{~k} \nn \\
&= 0~,
\eea
since $A_2\cdot B_1$ and $A_2\cdot B_2$ are simply numbers due to the fact that the gauge group $U(k_1)$ in \eref{quivSUN} is just $U(1)$ in this case; they hence commute with each other.  The equality labelled by ``$F$'' follows from the $F$-terms \eref{FtermsSUN}.  In addition, it is clear from the above analysis that
\bea \label{eeMM}
\epsilon^{i_1 \cdots i_N} \epsilon_{j_1 \cdots j_N} M^{j_1}_{~i_1} M^{j_2}_{~i_2} = 0
\eea

Thus, the reduced moduli space in question is indeed
\bea
\widetilde{\CM}_{1,\; SU(N),\; \CP^2} &=\{ M: \text{$M$ is an $N \times N$ matrix satisfying \eref{trMeq0}, \eref{M2eq0} and \eref{eeMM}} \}~, \nn \\
&= \widetilde{\CM}_{1,\; SU(N),\; \BC^2}~,
\eea
as expect from the second factor of \eref{HS1kSUN}.

\section{$USp(2N)$ instantons on $\CP^2$}\label{symplectic}
Let us now examine a construction and the moduli space of $USp(2N)$ instantons on $\CP^2$. Such an instanton bundle is characterised by their rank $N$ and their instanton number $\hat{k}$.

Elaborating on King's construction, the construction of $USp(2N)$ instantons on $\CP^2$ proposed by Bryan and Sanders (BS) in \cite{bryansanders2} can be realised from a $3d$ $\CN=2$ gauge theory with the quiver diagram \eref{ADHMC}, with the superpotential
\bea
W = \epsilon^{\alpha \beta} (S_\alpha)_{ab} \widetilde{S}^{bc} (S_\beta)_{cd} \widetilde{A}^{da}  + \widetilde{A}^{ab} Q^i_{~a} Q^j_{~b} J_{ij}~,
\eea
where the charges of the chiral fields are tabulated in \tref{tab:chargeADHMC}; $a,b,c,d=1,\ldots, k$ are the $U(k)$ gauge indices; $i,j=1, \ldots, 2N$ are the $USp(2N)$ flavour indices with $J_{ij}$ the corresponding symplectic matrix
\bea
J = \begin{pmatrix} 0 & {\bf 1}_{N\times N} \\ -{\bf 1}_{N\times N}  & 0\end{pmatrix}~,
\eea
and $\alpha, \beta=1,2$ are the indices corresponding to the $SU(2)$ global symmetry under which $S_1,\; S_2$ transform as a doublet.
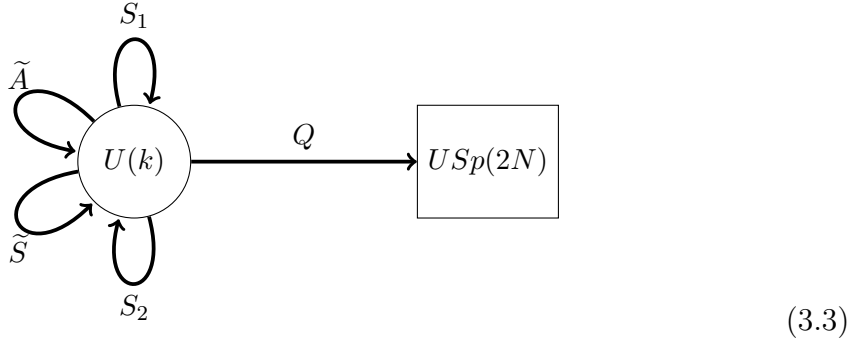
\bea \label{ADHMC}
\begin{tikzpicture}[font=\small]
\begin{scope}[auto,%
  every node/.style={draw, minimum size=1.5cm}, node distance=3cm];
\node[circle] (k) at (0, 0) {$U(k)$};
\node[rectangle, right=of k] (CN)  {$USp(2N)$};
\end{scope}
\draw[draw=black,solid,line width=0.5mm,->]    (k) to node[midway,above] {$Q$} (CN) ;
\draw[draw=black,solid,line width=0.5mm,->]    (k) edge [in=170,out=135,loop] node[above] {$\widetilde{A}$} (k) ;
\draw[draw=black,solid,line width=0.5mm,->]    (k) edge [in=225,out=190,loop] node[below] {$\widetilde{S}$} (k) ;
\draw[draw=black,solid,line width=0.5mm,<-]    (k) edge [loop above] node {$S_1$} (k) ;
\draw[draw=black,solid,line width=0.5mm,->]    (k) edge [loop below] node {$S_2$} (k) ;
\end{tikzpicture}
\eea
The CS level of the gauge group $U(k)$ can take any integral value\footnote{The cubic indices for the relevant representations of $SU(k)$ are 
\bea 
&d_3([1,0,\ldots,0])= 1, \quad d_3([0,1,0,\ldots,0]) =k-4, \quad d_3([2,0,\ldots,0]) = k+4, \; \nn \\
&d_3([0,\ldots,0,1,0]) =-(k-4),\quad d_3([0,\ldots,0,2]) = -(k+4)~. \nn
\eea}.
\begin{table}[htdp]
\begin{center}
\begin{tabular}{|c||c||c|c|c|}
\hline
~ & $U(k)$ & $USp(2N)$ & $SU(2)$ & $U(1)_R$ \\
\hline
$Q$ & $[1,0,\ldots,0]_{+1}$ & $[1,0, \ldots, 0]$ & $[0]$ & $1/2$ \\
$S_1, \; S_2$ & $[2,0, \ldots,0]_{+2}$ & $[\vec 0]$ & $[1]$ & $1/4$ \\
$\widetilde{A}$ & $[0,1,0, \ldots,0]_{-2}$ & $[\vec 0]$ & $[0]$ & $1$ \\
$\widetilde{S}$ & $[2,0, \ldots,0]_{-2}$ & $[\vec 0]$ & $[0]$ & $1/2$ \\
\hline
\end{tabular}
\end{center}
\caption{Charge assignments of each chiral field in the quiver diagram in \eref{ADHMC}.}
\label{tab:chargeADHMC}
\end{table}%

The branch of the moduli space that can be identified with that of $k$ $USp(2N)$ instantons on $\CP^2$, denoted by $\CM^{USp(2N)}_{\CP^2}$, is the one on which $ \widetilde{A}=0$.  The relevant $F$-terms are
\bea \label{FtermsC}
0=\frac{\partial W}{\partial \widetilde{A}^{da}} = \epsilon^{\alpha \beta} (S_\alpha)_{ab} \widetilde{S}^{bc} (S_\beta)_{cd}   + Q^i_{~d} Q^j_{~a} J_{ij}~.
\eea
This is indeed the integrability condition given in Table 1 of \cite{bryansanders2}.

The Hilbert series of the instanton branch of this theory is\footnote{The same comment as Footnote \ref{CIremark} applies here.}
\bea
& H[k, USp(N), \CP^2](t, x, \vec y) \nn \\
&= \int \ud \mu_{U(k)}(\vec z) \PE \Big[ \chi_{S_1, S_2} t+ \chi_{\widetilde{S}} t^2+ \chi_Q t^2 - \chi_{\text{$F$-terms}} t^4 \Big]~,
\eea
where the contributions from each chiral field are
\bea
\begin{array}{ll}
\chi_{S_1,S_2} = (x+x^{-1}) \sum_{1 \leq a \leq b \leq k} z_a z_b~, \qquad & \chi_{\widetilde{S}} = \sum_{1 \leq a \leq b \leq k} z_a^{-1} z_b^{-1}  \\
\chi_{Q} = \left[ \sum_{i=1}^N (y_i +y^{-1}_i) \right] \left( \sum_{a=1}^k z_a \right)~, \qquad & \chi_{\text{$F$-terms}}=\sum_{1 \leq a < b \leq k} z_a z_b~.
\end{array}
\eea
Indeed, as a result of the integrations, we see that
\bea
H[k, USp(N), \CP^2](t, x, \vec y) = H[k, USp(N), \BC^2](t^3, x, \vec y)~.
\eea

\subsubsection*{Mapping between the BS construction and the ADHM construction}

Similarly to the case of $U(N)$ instantons, it is possible to find a relation among the construction for instantons on $\CP^2$ and the ADHM construction for on $\BC^2$, whose quiver and the superpotential in the $4d$ $\CN=1$ notation are described in \eref{ADHMquivUSp}. 

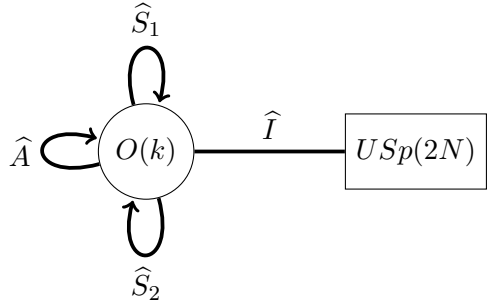
\bea \label{ADHMquivUSp}
& \hspace{0cm} \begin{tikzpicture}[font=\small]
\begin{scope}[auto,%
  every node/.style={draw, minimum size=1cm}, node distance=2cm];
\node[circle] (k) at (0,0) {$O(k)$};
\node[rectangle, right=of k] (N) {$USp(2N)$};
\end{scope}
\draw[draw=black,solid,line width=0.5mm,-] (k) to node[midway,above] {$\widehat{I}$}
      node[midway,below] {} (N);
\draw[draw=black,solid,line width=0.5mm,->]    (k) edge [loop above] node {$\widehat{S}_1$} (k) ;
\draw[draw=black,solid,line width=0.5mm,->]    (k) edge [loop left] node {$\widehat{A}$} (k) ;
\draw[draw=black,solid,line width=0.5mm,->]    (k) edge [loop below] node {$\widehat{S}_2$} (k) ;
\end{tikzpicture}
\nn \\
W_{\text{ADHM}} &= J_{ij}  \widehat{I}^i_{~a}  \widehat{A}_{a b} \widehat{I}^j_{~b}  + \epsilon^{\alpha \beta}  (\widehat{S}_\alpha)_{a b} \widehat{A}_{bc} (\widehat{S}_\beta)_{ca}~,
\eea

\noindent The corresponding $F$-terms of this ADHM quiver are
\bea
0=\partial_{\widehat{A}_{a'a}} W_{\text{ADHM}} &=  J_{ij} \widehat{I}^i_{~a'} \widehat{I}^j_{~a} + \epsilon^{\alpha \beta}  (\widehat{S}_\alpha)_{ba'} (\widehat{S}_\beta)_{ab}~.
\eea
By considering \cite{bryansanders2}
\bea
\sigma \,:\, (\widehat{S}_1, \widehat{S}_2, \widehat{I}) ~ \rightarrow ~ \left(S_1 = \widehat{S}_{1} , \; S_2 = \widehat{S}_{2} , \;  \widetilde{S} =  {\bf 1}_{k \times k} ,\;  Q =\widehat{I} \right )\, ,
\eea
one can see that $\sigma$ maps \eref{ADHMquivUSp} to \eref{FtermsC} upon multiplication by $\widetilde{S}^{a'd}$. This map descends to a pull-back of the moduli spaces in agreement with our finding at the level of the Hilbert series computation. 

One might be tempted to construct in addition the ``inverse map"
\bea
&\pi\, :\, (S_1, S_2, \widetilde{S}, Q) ~ \rightarrow ~ (\widehat{S}_{1}, \; \widehat{S}_{2} ,\; \widehat{I} )~ , \nn \\
&\widehat{I}^i_{~a}\widehat{I}^j_{~b}=\widetilde{S}^{ab'} Q^i_{~b'}  Q^j_{~b}~, \quad (\widehat{S}_\alpha)_{ab} =  (S_\alpha)_{ab'} \widetilde{S}^{b'b}~.
\eea
Here $\widehat{S}_{\alpha}$ (with $\alpha=1,2$) are the rank-two symmetric chiral fields, $\widehat{A}$ is the rank-two anti-symmetric chiral fields coming from the $4d$ $\CN=2$ vector multiplet of the $O(k)$ gauge group and $Q$ is the bifundamental chiral field. We emphasise that the group under which the indices transform has been changed from those of $SO(k)$ to $U(k)$.  However, the equation defining the $\hat{I}$ is quadratic --that is, it is of the form $\hat{I}^2=\tilde{S}\, Q^2$--, as opposed to the unitary case. Hence, even though $\pi$ converts the $\CP^2$ $F$-terms into the $S^4$ $F$-terms, $\pi$ is not an algebraic mapping.

\subsubsection*{Properties of the instanton moduli space}

It is natural to guess that, by allowing a different FI parameter associated with the gauge group $U(k)$ in the quiver \eref{ADHMC}, a resolved moduli space $\widehat{\mathcal{M}}^{USp(2\,N)}_{\CP^2}$ might exist also in this case. Its dimension would be computed in a similar way to \eref{diffdim}.

In this case, the number of degrees of freedom of the chiral fields $\{Q, S_1, S_2, \widetilde{S} \}$ is $2Nk+3\frac{k(k+1)}{2}$, while we have $\frac{k\,(k-1)}{2}$ $D$-term constraints and $k^2$ $F$-term constraints. Hence, the complex dimension of the resolved instanton moduli space is
\begin{equation}
{\rm dim}_{\mathbb{C}}\,\widehat{\mathcal{M}}^{USp(2\,N)}_{\CP^2} 
= 2Nk+3\frac{k(k+1)}{2} -\frac{k\,(k-1)}{2}-k^2 
=2k(N+1)~, \label{dimresolvedUSp}
\end{equation}
equal to to the dimension of the moduli space for $USp(2\,N)$ instantons on $\mathbb{C}^2$, which is in agreement with the result from the Hilbert series:
\bea
\dim \widehat{\mathcal{M}}^{USp(2\,N)}_{\CP^2}= \dim \mathcal{M}^{USp(2\,N)}_{\CP^2} = \dim \mathcal{M}^{USp(2\,N)}_{\BC^2}
\eea
From \eref{dimresolvedUSp}, the instanton number can be identified with the rank $k$ of the gauge symmetry $U(k)$ in the quiver \eref{ADHMC}.  Indeed, it can take only be integral, on the contrary to $SU(N)$ instantons on $\CP^2$. This, together with the fact that in this case there is no other class characterizing the instanton and potentially standing for extra dimensions in the moduli space, suggests that in this case  $\widehat{\mathcal{M}}^{USp(2\,N)}_{\CP^2}$ and $\mathcal{M}^{USp(2\,N)}_{\BC^2}$ might be directly related one to the other. In particular, this suggests that the maps $\pi$, $\sigma$ indeed provide an algebraic identification of the spaces.

\section{$SO(N)$ instantons on $\CP^2$}\label{orthogonal}
The construction of $SO(N)$ instantons on $\CP^2$ was provided in \cite{bryansanders2}. Being a real bundle, it is characterised by the rank $N$ and the instanton number $k$, and the second Stiefel-Whitney class.

The construction of $SO(N)$ instantons on $\CP^2$  provided in \cite{bryansanders2} can be realised from a $3d$ $\CN=2$ gauge theory with the quiver diagram \eref{ADHMSO}, with the superpotential
\bea
W = \epsilon^{\alpha \beta} (A_\alpha)_{ab} \widetilde{A}^{bc} (A_\beta)_{cd} \widetilde{S}^{da}  + \widetilde{S}^{ab} Q^i_{~a} Q^j_{~b} M^{SO(N)}_{ij}~,
\eea
where the charges of the chiral fields are tabulated in \tref{tab:chargeADHMSO}; $a,b,c,d=1,\ldots, 2k$ are the $U(2k)$ gauge indices with $k$ being {\it integral} or {\it half-odd-integral}; and $i,j=1, \ldots, N$ are the $SO(N)$ flavours indices with $M^{SO(N)}$ given by
\bea
M^{SO(2N)} = \begin{pmatrix} 0 & {\bf 1}_{N\times N} \\ {\bf 1}_{N\times N} & 0 \end{pmatrix}~, \qquad M^{SO(2N+1)} = \begin{pmatrix} 0 & {\bf 1}_{N\times N} & 0 \\ {\bf 1}_{N\times N} & 0 & 0 \\ 0 & 0 & 1 \end{pmatrix}~.
\eea
and $\alpha, \beta=1,2$ are the indices corresponding to the $SU(2)$ global symmetry under which $A_1,\; A_2$ transform as a doublet.
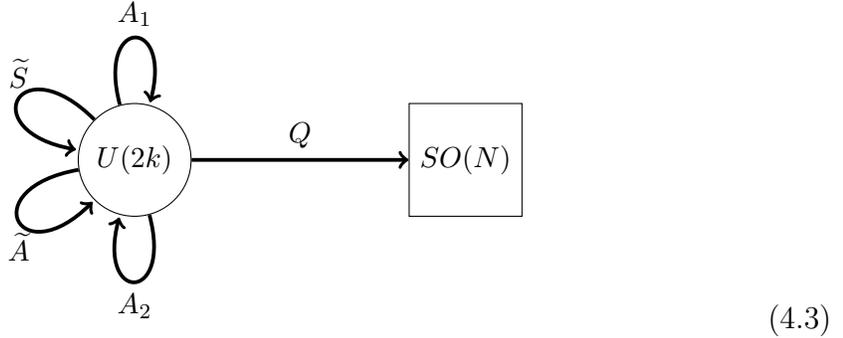
\bea \label{ADHMSO}
\begin{tikzpicture}[font=\small]
\begin{scope}[auto,%
  every node/.style={draw, minimum size=1.5cm}, node distance=3cm];
\node[circle] (2k) at (0, 0) {$U(2k)$};
\node[rectangle, right=of k] (SON)  {$SO(N)$};
\end{scope}
\draw[draw=black,solid,line width=0.5mm,->]    (2k) to node[midway,above] {$Q$} (SON) ;
\draw[draw=black,solid,line width=0.5mm,->]    (2k) edge [in=170,out=135,loop] node[above] {$\widetilde{S}$} (k) ;
\draw[draw=black,solid,line width=0.5mm,->]    (2k) edge [in=225,out=190,loop] node[below] {$\widetilde{A}$} (k) ;
\draw[draw=black,solid,line width=0.5mm,<-]    (2k) edge [loop above] node {$A_1$} (2k) ;
\draw[draw=black,solid,line width=0.5mm,->]    (2k) edge [loop below] node {$A_2$} (2k) ;
\end{tikzpicture}
\eea
The CS level for the $U(2k)$ gauge group can take the value $-N/2 + \kappa$, with $\kappa$ an integer.
\begin{table}[htdp]
\begin{center}
\begin{tabular}{|c||c||c|c|c|}
\hline
~ & $U(2k)$ & $SO(N)$ & $SU(2)$ & $U(1)_R$ \\
\hline
$Q$ & $[1,0,\ldots,0]_{+1}$ & $[1,0, \ldots, 0]$ & $[0]$ & $1/2$ \\
$A_1, \; A_2$ & $[2,0, \ldots,0]_{+2}$ & $[\vec 0]$ & $[1]$ & $1/4$ \\
$\widetilde{A}$ & $[0,1,0, \ldots,0]_{-2}$ & $[\vec 0]$ & $[0]$ & $1/2$ \\
$\widetilde{S}$ & $[2,0, \ldots,0]_{-2}$ & $[\vec 0]$ & $[0]$ & $1$ \\
\hline
\end{tabular}
\end{center}
\caption{Charge assignments of each chiral field in the quiver diagram in \eref{ADHMSO}. Note that $k$ can be integral or half-odd-integral.}
\label{tab:chargeADHMSO}
\end{table}%

The branch of the moduli space that can be identified with that of $k$ $SO(N)$ instantons on $\CP^2$, denoted by $\CM^{SO(N)}_{\CP^2}$, is the one on which $ \widetilde{S}=0$.  The relevant $F$-terms are
\bea \label{FtermsSO}
0=\partial_{\widetilde{S}^{da}} W= \epsilon^{\alpha \beta} (A_\alpha)_{ab} \widetilde{A}^{bc} (A_\beta)_{cd}   + Q^i_{~d} Q^j_{~a} M^{SO(N)}_{ij}~.
\eea
This is indeed the integrability condition given in Table 1 of \cite{bryansanders2}.

The Hilbert series of the instanton branch of this theory is\footnote{The same comment as Footnote \ref{CIremark} applies here for $N\geq 5$.}
\bea
& H[k, SO(N), \CP^2](t, x, \vec y) \nn \\
&= \int \ud \mu_{U(2k)}(\vec z) \PE \Big[  \chi_{A_1, A_2}t + \chi_{\widetilde{A}} t^2+ \chi_Q t^2 - \chi_{\text{$F$-terms}} t^4 \Big]~,
\eea
where the contributions from each chiral field are
\bea
\begin{array}{ll}
 \chi_{Q} =\left( \sum_{a=1}^{2k} z_a \right) \times 
\begin{cases} 
\sum_{i=1}^{N/2} (y_i +y^{-1}_i)~, &\qquad  \text{$N$ even}\\
\sum_{i=1}^{(N-1)/2} (y_i +1 +y^{-1}_i)~,&\qquad  \text{$N$ odd}~,
\end{cases} & ~ \\
 \chi_{\widetilde{A}} = \sum_{1 \leq a < b \leq 2k} z_a^{-1} z_b^{-1}~, \\
\chi_{A_1,A_2} = (x+x^{-1}) \sum_{1 \leq a < b \leq 2k} z_a z_b~, \\
 \chi_{\text{$F$-terms}} =\sum_{1 \leq a \leq b \leq 2k} z_a z_b~.
\end{array}
\eea
Indeed, as a result of the integrations, we see that
\bea
H[ k , SO(N), \CP^2](t, x, \vec y) 
&= H[\lfloor k \rfloor, SO(N), \BC^2](t^3, x, \vec y)~,
\eea
where $\lfloor k \rfloor$ denotes the floor function selecting the largest integer that is smaller or equal to $k$. It then follows that 
\begin{equation}
\dim_\BC\,\mathcal{M}^{SO(N)}_{\CP^2}=2(N-2)\lfloor k \rfloor\ .
\end{equation}
\subsubsection*{Mapping between the BS construction and the ADHM construction}

As in the previous cases, it is possible to relate the construction for instantons on $\CP^2$ into the ADHM construction for on $\BC^2$, whose quiver and the superpotential in the $4d$ $\CN=1$ notation are described in \eref{ADHMquivSO}.

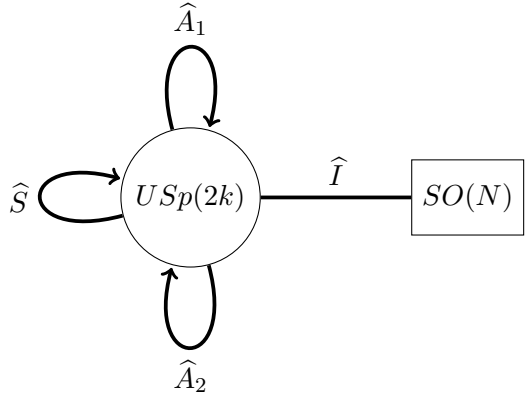
\bea \label{ADHMquivSO}
& \hspace{0cm} \begin{tikzpicture}[font=\small]
\begin{scope}[auto,%
  every node/.style={draw, minimum size=1cm}, node distance=2cm];
\node[circle] (k) at (0,0) {$USp(2k)$};
\node[rectangle, right=of k] (N) {$SO(N)$};
\end{scope}
\draw[draw=black,solid,line width=0.5mm,-] (k) to node[midway,above] {$\widehat{I}$}
      node[midway,below] {} (N);
\draw[draw=black,solid,line width=0.5mm,->]    (k) edge [loop above] node {$\widehat{A}_1$} (k) ;
\draw[draw=black,solid,line width=0.5mm,->]    (k) edge [loop left] node {$\widehat{S}$} (k) ;
\draw[draw=black,solid,line width=0.5mm,->]    (k) edge [loop below] node {$\widehat{A}_2$} (k) ;
\end{tikzpicture}
\nn \\
W_{\text{ADHM}} &= J^{ba'} J^{ac} \widehat{I}^i_{~b}  \widehat{S}_{a'a} \widehat{I}^j_{~c} M^{SO(N)}_{ij}  + \epsilon^{\alpha \beta}  J^{cc'} J^{ba'} J^{ab'} (\widehat{A}_\alpha)_{c b} \widehat{S}_{a'a} (\widehat{A}_\beta)_{b'c'}~,
\eea

\noindent The corresponding $F$ terms are now
\bea
0=\partial_{\widehat{S}_{a'a}} W_{\text{ADHM}} &=  J^{ba'} J^{ac} \widehat{I}^i_{~b} \widehat{I}^j_{~c} M^{SO(N)}_{ij} + \epsilon^{\alpha \beta}  J^{a'b} (\widehat{A}_\alpha)_{bc} J^{cc'} (\widehat{A}_\beta)_{c'b'} J^{b'a}~,
\eea

\noindent Defining \cite{bryansanders2}
\bea
\sigma \,:\, (\widehat{A}_1, \widehat{A}_2, \widehat{I}) ~ \rightarrow ~ \left((A_\alpha)_{ab} = J^{ab'}(\widehat{A}_{\alpha})_{b'b} , \;  \widetilde{A}^{ab} =  J_{ab} ,\;  Q^i_{~a} =J^{ab} \widehat{I}^{i}_{~b} \right )\, ,
\eea
It is straightforward to see that $\sigma$ maps \eref{ADHMquivUSp} to \eref{FtermsC} after multiplication by $\widetilde{A}^{da'}$, hence finding again a pull-back of moduli spaces consistent with our computation.

Exactly as in the symplectic case, one might consider an ``inverse map"
\bea
&\pi\, :\, (A_1, A_2, \widetilde{A}, Q) ~ \rightarrow ~ (\widehat{A}_{1}, \; \widehat{A}_{2} ,\; \widehat{I} )~ , \quad \text{such that}\nn \\
&\widetilde{A}^{da'} Q^i_{~d} Q^i_{~a} = J^{ba'} J^{ac} \widehat{I}^i_{~b} \widehat{I}^j_{~c}~, \quad (A_\alpha)_{ab'}\widetilde{A}^{b'a'} = (\widehat{A}_\alpha)_{ab} J^{ba'} ~.
\eea
where $\widehat{A}_{\alpha}$ (with $\alpha=1,2$) are the rank-two symmetric chiral fields, $\widehat{S}$ is the rank-two anti-symmetric chiral fields coming from the $4d$ $\CN=2$ vector multiplet of the $USp(2k)$ gauge group and $Q$ is the bifundamental chiral field.  We emphasise that the group under which the indices transform has been changed from those of $U(2k)$ to $USp(2k)$. 

This relation is, as in the symplectic case, quadratic, and hence does not define an algebraic map. Nevertheless direct use of it converts the  $\CP^2$ $F$-terms into the $S^4$ $F$-terms.

\subsubsection*{Properties of the instanton moduli space}

Let us introduce the resolved moduli space $\widehat{\mathcal{M}}^{SO(N)}_{\CP^2}$ of $SO(N)$ instantons on $\CP^2$ by allowing a different FI parameter associated with the gauge group $U(2k)$ in the quiver \eref{ADHMSO}.  The dimension of the resolved moduli space can be computed in a similar way to \eref{diffdim}.

In this case the total degrees of freedom of $\{Q,A_1,A_2, \widetilde{A}\}$ are $2Nk+3k(2\,k-1)$, while the $F$-terms impose $k(2k+1)$ conditions and the $D$-terms inpose $4k^2$ conditions. Thus, the expected complex dimension of the resolved moduli space is
\begin{equation} \label{resolvedSON}
{\rm dim}_{\mathbb{C}}\,\widehat{\mathcal{M}}^{SO(N)}_{\CP^2}= 2Nk+3k(2k-1) - k(2k+1) - 4k^2=2k(N-2)~.
\end{equation}
We therefore identify $k$ as an instanton number and emphasise that $k$ can be either an integer or a half-odd-integer. On the other hand, as we have shown above, at least at the algebraic level, the moduli space of $k$ $SO(N)$ instantons on $\CP^2$ can be mapped to that of $\lfloor k \rfloor$ $SO(N)$ instantons on $\BC^2$. It is then natural to identify $2(k-\lfloor k\rfloor)$ with the second Stiefel-Whitney class; hence, if $k$ is an integer the second Stiefel-Whitney class is trivial and if $k$ is a half-odd-integer it is non-trivial. Furthermore, analogously to the role of the first Chern class $\hat{c}$ for the unitary instantons, the second Stiefel-Whitney class is responsible for the extra directions that are present in  $\widehat{\mathcal{M}}^{SO(N)}_{\CP^2}$ but not ${\mathcal{M}}^{SO(N)}_{\CP^2}$.  We again expect that these extra directions are related to the moduli space of certain dissolved surface operators. Presumably, the quadratic nature of the map $\pi$ might be related to these issues.

It is interesting to analyze in detail the case of $SO(6)$, whose double cover is $SU(4)$. The obstruction to lift an $SU(4)$ bundle to $SO(6)$ is represented by the second Stiefel-Whitney class, which can be seen as the reduction modulo 2 of the first Chern class. Thus, an $SU(4)$ instanton bundles can be identified with an $SO(6)$ bundle if and only if $\hat{c}=0$ modulo $2$. Indeed, it can be seen from \eref{kkhat}, \eref{diffdim} and \eref{resolvedSON} that the dimension of the resolved space of $SU(4)$ instantons can be equated to that of $SO(6)$ instantons if and only if $\hat{c}$ vanishes modulo 2.

It is also very interesting to consider the $SO(3)$ case. However in this case the gauge symmetry of the ADHM quiver is not fully broken, hence requiring the aid of computing programs such as {\tt Macaulay2} \cite{M2} to compute the Hilbert series. Unfortunately the computing needs increase very rapidly with the instanton number and we have not been able to perform detailed tests. Note, however, that the spin group for $SO(3)$ is $SU(2)$, for which the first Chern class mod 2 vanishes only for $\hat{c}=0$. This would suggest that an $SO(3)$ instanton bundle with the instanton number $k$ can be identified with an $SU(2)$ bundle with the instanton number $2k$ and the vanishing first Chern class.

\section{The blowup formula}\label{blowupformula}

In this section, we present another method to compute the Hilbert series of instantons on $\CP^2$, namely \textit{via} the blowup formula. Recall that framed SD connections on $\CP^2$ (or, equivalently, framed ASD connections on $\overline{\CP}^2$) constructed through the King construction are in one-to-one correspondence with holomorphic bundles\footnote{Strictly speaking the should talk about torsion free sheaves.} on $\mathbb{F}_1$ with trivial behavior at $\ell_{\infty}$. Hence, we can interchangeably refer to sheaves on $\mathbb{F}_1$ or ASD bundles (\textit{i.e.} holomorphic bundles with ASD connection) on $\overline{\CP}^2$. The blowup formula in fact precisely constructs those bundles. 

For instantons in a special unitary gauge group, such a formula for the blowup of $\CP^2$ was given in (2.2) of \cite{Nakajima:2005fg}.  Moreover, since $\CP^2$ blown up at one point can be identified with the Hirzebruch surface $\BF_1$, a similar blowup formula can be obtained from (3.21) of \cite{Bonelli:2012ny}, with the surface $X_{p,q}$ being $X_{1,1} = \BF_1$.  A generalisation for instantons in a general gauge group was discussed in \cite{Nakajima:2003pg, Keller:2012da}.

Let us summarise the computation using the blowup formula.  First of all, we define the generating function $\CH[G, \BC^2]$ for the Hilbert series of $G$ instantons on $\BC^2$ as
\bea
\CH[G, \BC^2] (t_1,t_2; \vec z; q) = \sum_{k=0}^\infty H[k, G, \BC^2](t_1,t_2; \vec z)  q^k~,
\eea
where $H[k, G, \BC^2](t_1,t_2; \vec z)$ denotes the Hilbert series of $k$ $G$ instantons on $\BC^2$ and $q$ is the fugacity associated with the instanton number for instantons on $\BC^2$.  Then, the blowup formula for a one-point-blown-up of $\CP^2$, \ie~ the Hirzebruch surface $\BF_1$, reads
\bea \label{blowup}
&\CH[G; \hat{c}; \BF_1] (t_1,t_2; \vec z; q) \nn \\
&= \sum_{\vec v \in \CV_G(\hat{c})} \frac{q^{\frac{1}{2} \vec v \cdot \vec v}}{\prod_{\vec \alpha \in \Delta_G} \ell^{\BF_1}_{\vec \alpha} (t_1, t_2; \vec z; \vec v)}\prod_{r=0}^1 \CH[G, \BC^2](t_1^{(r)},t_2^{(r)}; \vec z^{(r)}; q)~,
\eea
where $\CH[G; \hat{c}; \BF_1](t_1,t_2; \vec z; q)$ is the generating function for the Hilbert series of instantons on $\CP^2$, and $r=0, \;1$ labels the `patches', each of which contains the Hilbert series of instantons on $\BC^2$.   It should be emphasised that even though we indicate $\BF_1$ in the square bracket to indicate the blown-up space, the space on which instantons live is $\CP^2$.

For $G$ a unitary group, the parameter $\hat{c}$ has an interpretation of the first Chern class of the instanton bundle.  In general, we observe that $\hat{c}$ is related with the homotopy group $\pi_1(G)$, as explained below.  The notations in the \eref{blowup} are defined as follows.
\bi
\item The parameters $t_1^{(r)}$,$t_2^{(r)}$ and $\vec z^{(r)}$, with $r=0,1$, are defined as follows:
\bea
\begin{array}{lll}
t_1^{(0)} = t_1~, &\qquad  t_2^{(0)} = t_2 t_1^{-1}~, &\qquad z_a^{(0)}= t_1^{v_a} z_a~, \\
t_1^{(1)} = t_1 t_2^{-1}~, &\qquad t_2^{(1)} = t_2~, &\qquad z_a^{(1)} = t_2^{v_a} z_a~.
\end{array}
\eea
\item We denote the set of roots of $G$ by $\Delta_G$.  For any root $\vec \alpha \in \Delta_G$,  the corresponding coroot $\alpha^\vee$ of $G$ is defined as $\vec \alpha^\vee = 2 \vec \alpha/ (\vec \alpha \cdot \vec \alpha)$. The normalisation for the root system is such that the squared length of the long roots is $2$.  We also take ``$\cdot$'' to be an inner product with respect to the standard orthonormal basis $\{ \vec e_i \}$.
\item The set $\CV_G(c)$ is a set of the fixed points for the blowup formula.  For each group $G$, it is given as follows.
\ben
\item For a unitary group or strictly speaking $G =SU(N)/\BZ_N$, $\pi_1(G) =\BZ_N$; hence, $\hat{c}$ is an integer modulo $N$.
The set $\CV_G(\hat{c})$ is defined as

\bea \label{setVSUN}
\CV_{SU(N)} (\hat{c}) = \left \{  \vec v \in \BQ^{\rk(G)} \Big | \sum_a v_a = 0, ~v_a = -\frac{\hat{c}}{N}~(\rm{mod} \; \BZ), ~\text{$1 \leq a \leq N$} \right \}~.\nn
\eea
Note that, for $\hat{c}=0$, the set $\CV_{SU(N)} (\hat{c}=0)$ is spanned by the simple roots of $SU(N)$, namely 
\bea \vec e_1-\vec e_2, \; \vec e_2- \vec e_3, \; \ldots, \;  \vec e_{N-1}- \vec e_{N}~. \eea
In the language of the King quiver \eref{quivSUN}, $\hat{c} = |k_1-k_2|$.  Note that the powers of $q$ appearing in the blowup formula \eref{blowup} indicate the instanton numbers $\hat{k}$ given by \eref{kkhat}.
\item For $G=SO(N)$, the homotopy group $\pi_1(G) =\BZ_2$.  In this case, the parameter $\hat{c}$, which can be $0$ or $1$, has an interpretation of the Stiefel-Whitney class. 
The sets $\CV_{G}$, for $G=B_N=SO(2N+1)$ and $G=D_N=SO(2N)$, are given by
\bea
\CV_{B_N, D_N} (\hat{c}) = \left \{  \vec v \in \BZ^{N} \Big | \sum_{a=1}^N v_a =\hat{c} = 0, 1~ (\mathrm{mod}\; 2) \right \}~. \label{setVBNCN}
\eea
Indeed, for $\hat{c}=0$,  the sets $\CV_{B_N, C_N} (\hat{c}=0)$ are spanned by their simple roots; respectively, these are
\bea
G=B_N: \quad & \vec e_1-\vec e_2, \; \vec e_2- \vec e_3, \; \ldots, \;  \vec e_{N-1}- \vec e_{N}, \; \vec e_N~. \\
G=D_N: \quad & \vec e_1-\vec e_2, \; \vec e_2- \vec e_3, \; \ldots, \;  \vec e_{N-1}- \vec e_{N}, \; \vec e_{N-1}+ \vec e_N~.
\eea
The powers of $q$ in \eref{blowup} indicates the instanton numbers.  Such numbers are integral if $\hat{c}=0$ and half-odd-integral if $\hat{c}=1$.
\item For $G= USp(2N)$, the homotopy group $\pi_1(G)$ is trivial; hence, the only possible value of $\hat{c}$ is zero.  The set $\CV_{USp(N)}$ is spanned by the simple roots of $USp(N)$, namely 
\bea \frac{1}{\sqrt{2}}   ( \vec e_1-\vec e_2), \; \frac{1}{\sqrt{2}} (\vec e_2- \vec e_3), \; \ldots, \; \frac{1}{\sqrt{2}}( \vec e_{N-1}- \vec e_{N}), \; \sqrt{2} \vec e_{N}~. \eea
In this case and below, the powers of $q$ appearing in \eref{blowup} correspond to the instanton numbers.
\item For $G=G_2$, the homotopy group $\pi_1(G)$ is trivial; the only possible value of $\hat{c}$ is zero. The set $\CV_{G_2}$ is spanned by the simple roots
\bea
\sqrt{\frac{2}{3}} \vec e_1, \quad  -\sqrt{\frac{3}{2}} \vec e_1 + \frac{1}{\sqrt{2}} \vec e_2~. 
\eea
\item For $G=F_4, \; E_6, \; E_7, \; E_8$, their $\pi_1(G)$ are trivial, and so we expect $\CV_{G}$ to be spanned by their respective simple roots with the appropriate normalisation.  Although explicit checks for these groups are possible, they can be rather cumbersome in practice and we shall not present any computation for such groups in this paper.
\een
\item The factor $ \ell^{\BF_1}_{\vec \alpha} (t_1, t_2; \vec z; \vec v)$ is defined as
\bea
\ell^{\BF_1}_{\vec \alpha} (t_1, t_2; \vec z; \vec v) = \begin{cases} \prod \limits_{\substack{i, j \geq 0 \\ i+j \leq -\vec (\vec v \cdot \vec \alpha)-1}} \left( 1- t_1^{-i} t_2^{-j} \prod_{a=1}^{\rk(G)} z_a^{\alpha_a} \right)  & \quad \text{for}~\vec v \cdot \vec \alpha < 0~, \\ 
 \prod \limits_{\substack{i, j \geq 0 \\ i+j \leq \vec (\vec v \cdot \vec \alpha)-2}} \left( 1- t_1^{i+1} t_2^{j+1} \prod_{a=1}^{\rk(G)} z_a^{\alpha_a} \right)  & \quad \text{for}~\vec v \cdot \vec \alpha > 1~, \\
1  & \quad \text{otherwise~.}
\end{cases}
\eea
\item The explicit expressions for the Hilbert series of one and two $G$ instantons on $\BC^2$, with $G$ a simple group, are given in \cite{Benvenuti:2010pq,Hanany:2012dm}.  For one instanton, the formula takes a simple expression:
\bea
H[k=1; G; \BC^2](t_1, t_2; \vec z) = \frac{1}{(1-t_1)(1-t_2)} \sum_{p=0}^\infty  \chi^{G}_{p \cdot {\bf Adj}} (\vec z) t^{2p}~,
\eea
where $p \cdot {\bf Adj}$ denotes the irreducible representation of $G$ whose highest weight is equal to $p$ times that of the adjoint representation, and $\chi^{G}_{p \cdot {\bf Adj}} (\vec z)$ denotes its character written in terms of $\vec z$.  For $G=SU(N)$, explicit expressions of Hilbert series for higher instanton numbers $k$ can be efficiently computed by summing over contributions over partitions of Young diagrams \cite{Flume:2002az, Nekrasov:2002qd, Bruzzo:2002xf, Nakajima:2003pg}; see also section 2.3.1 of \cite{Dey:2013fea}.
\ei

One interesting observation that applies to all simple groups $G$ of our interest is as follows. When $\hat{c}$ is an even number, the powers of $q$ in \eref{blowup} are integral; this corresponds to the integer-valued instanton numbers.  On the other hand, if $\hat{c}$ is odd, the powers of $q$ in \eref{blowup} are half-odd-integral, corresponding to half-odd-integer-valued instanton numbers.

Below we demonstrate the use of blowup formula \eref{blowup} in various examples, including instantons in $A$, $B$, $C$, $D$ and $G$ type gauge groups.

\subsection{$SU(2)$ instantons}
In this subsection, we take $G=SU(2)$.  The set of roots is
\bea
\Delta_{SU(2)} = \{ (1,-1), (-1,1) \}~.
\eea
Below we demonstrate the blowup formula \eref{blowup} for certain small values of the first Chern classes.

\subsubsection{The first Chern class $\hat{c}=0$}
From \eref{setVSUN}, the set $\CV(\hat{c}=0)$ is given by
\bea
 \CV(\hat{c}=0) =  \{ (m,-m) | m \in \BZ  \}
\eea
From \eref{blowup}, we see that the powers of $q$ are $0, \; 1, \; 2, \; \ldots$, corresponding to the instanton numbers.  According to Theorem 2.4 of \cite{Nakajima:2005fg}, the Hilbert series for $SU(2)$ instantons on $\BF_1$ with the vanishing first class $\hat{c}=0$ and the instanton number $k$ is equal to that of $k$ $SU(2)$ instantons on $\BC^2$.   We demonstrate this statement for a few small instanton numbers below.

\subsubsection*{The instanton number $\hat{k}=1$}
This configuration of the instanton can be realised using the King quiver \eref{quivSUN}, with $k_1=k_2=1$ and $N=2$.  Below we compute the Hilbert series using the blowup formula.

The contributions to coefficient of $q^{1/2}$ in \eref{blowup} come from 
\bea \vec v =(1,-1), \; (0,0), \; (-1,1)~. \eea
From the blowup formula \eref{blowup}, we can extract the coefficient $q$, which gives the Hilbert series for the $SU(2)$ instanton with $\hat{c}=0$ and instanton number $1$, as
\bea \label{Hc2oneSU2}
&H[kk=1; SU(2); \hat{c}=0; \BF_1] (t_1,t_2,\vec z) \nn \\
&= \frac{1}{\prod_{\vec \alpha \in \Delta_{SU(2)}} \ell^{\BF_1}_{\vec \alpha} (t_1, t_2; \vec z; (0,0))} \left[ \sum_{r=0}^1 H[k=1; SU(2); \BC^2](t_1^{(r)},t_2^{(r)}; \vec z^{(r)}) \right]_{\vec v=(0,0)}  \nn \\
& \quad + \frac{1}{\prod_{\vec \alpha \in \Delta_{SU(2)}} \ell^{\BF_1}_{\vec \alpha} (t_1, t_2; \vec z; (1,-1))}  +\frac{1}{\prod_{\vec \alpha \in \Delta_{SU(2)}} \ell^{\BF_1}_{\vec \alpha} (t_1, t_2; \vec z; (-1,1))} \nn \\
&= H[k=1; SU(2); \BC^2] (t_1, t_2; \vec z)~,
\eea
where the Hilbert series of $1$ $SU(2)$ instanton on $\BC^2$ is
\bea \label{HS1SU2C2}
&H[k=1; SU(2); \BC^2] (t_1, t_2; \vec z) \nn \\
&=  \PE [t_1+t_2] \times \PE \left[ t_1 t_2 (1+z_1 z_2^{-1}+z_2 z_1^{-1}) - (t_1 t_2)^2 \right] \nn \\
&= \frac{1}{(1-t_1)(1-t_2)} \times \frac{1-(t_1 t_2)^2}{(1-t_1 t_2)(1-t_1 t_2 z_1 z_2^{-1})(1-t_1 t_2 z_1^{-1} z_2)}~,
\eea
and various $\ell$-factors are given by
\bea
 \ell^{\BF_1}_{\vec \alpha} (t_1, t_2; \vec z; (0,0)) &= 1~, \qquad \text{for all $\vec \alpha \in \Delta_{SU(2)}$}~, \nn \\
\ell^{\BF_1}_{(1,-1)} (t_1, t_2; \vec z; (1,-1)) &= 1- t_1 t_2 \frac{z_1}{z_2}~, \nn \\
\ell^{\BF_1}_{(-1,1)} (t_1, t_2; \vec z; (1,-1)) &= \left(1-\frac{z_2}{z_1}\right) \left(1-\frac{z_2}{t_1 z_1}\right) \left(1-\frac{z_2}{t_2 z_1}\right)~, \nn \\
\ell^{\BF_1}_{\vec \alpha} (t_1, t_2; (z_1,z_2) ; (-1,1)) &= \ell^{\BF_1}_{-\vec \alpha} (t_1, t_2; (z_2,z_1); (1,-1))~.
\eea

\subsubsection*{The instanton number $\hat{k}=2$}
Using the blowup formula \eref{blowup} and extracting the coefficient of $q^2$, the Hilbert series of question is given by
\bea
&H[k=2; SU(2); \hat{c}=0; \BF_1] (t_1,t_2,\vec z) \nn \\
& =\frac{1}{\prod_{\vec \alpha \in \Delta_{SU(2)}} \ell^{\BF_1}_{\vec \alpha} (t_1, t_2; \vec z; (0,0))}  \Bigg[ \sum_{r=0}^1 H[k=2; SU(2); \BC^2](t_1^{(r)},t_2^{(r)}; \vec z^{(r)}) +\nn \\
& \qquad \prod_{s=0}^2 H[k=1; SU(2); \BC^2](t_1^{(s)},t_2^{(s)}; \vec z^{(s)}) \Bigg]_{\vec v=(0,0)} +  \nn \\
& \quad \sum_{\vec v =\pm(1,-1)}  \frac{1}{\prod_{\vec \alpha \in \Delta_{SU(2)}} \ell^{\BF_1}_{\vec \alpha} (t_1, t_2; \vec z; \vec v)}  \Bigg[ \sum_{r=0}^1 H[k=1; SU(2); \BC^2](t_1^{(r)},t_2^{(r)}; \vec z^{(r)}) \Bigg]_{\vec v} \nn \\
&=  H[k=2; SU(2); \BC^2](t_1,t_2,\vec z)~.
\eea
where the above $\ell$-factors are unity, and the Hilbert series for two $SU(2)$ instantons $H[k=2; SU(2); \BC^2](t_1,t_2,\vec z)$ is given by Eq. (3.12) of \cite{Hanany:2012dm}.

Note that this configuration of instantons can be realised from quiver \eref{quivSUN} with $k_1=k_2=2$ and $N=2$.

\subsubsection{The first Chern class $\hat{c}=1$}
From \eref{setVSUN}, the set $\CV(\hat{c}=1)$ is given by
\bea
 \CV(\hat{c}=1) =  \{ (m+1/2,-m-1/2) | \; m \in \BZ  \}~.
\eea
From \eref{blowup}, we see that the smallest power of $q$ for $\hat{c}=1$ is $1/4$, corresponding to the instanton number $\hat{k}= 1/4$.  The next higher power of $q$ is $3/4$.  We examine these two cases below.

\subsubsection*{The instanton number $\hat{k} =1/4$}
The contributions to coefficient of $q^{1/4}$ in \eref{blowup} come from 
\bea \vec v =(1/2,-1/2), \; (-1/2,1/2) \eea 
in the product of the $\ell$-factors, and not from the non-trivial parts of $\CH[G, \BC^2](t_1,t_2; \vec z; q)$, since the latter contains only positive powers of $q$.  The relevant $\ell$-factors are given by
{\small
\bea \label{Lfacc1}
\begin{array}{ll}
\ell^{\BF_1}_{(-1,1)} (t_1, t_2; \vec z; (1/2,-1/2)) = 1-\frac{z_2}{z_1}~, &\quad \ell^{\BF_1}_{(1,-1)} (t_1, t_2; \vec z; (-1/2,1/2) ) =1 -\frac{z_1}{z_2}~, \\
\ell^{\BF_1}_{(1,-1)} (t_1, t_2; \vec z; (-1/2,1/2) ) =1~,&\quad \ell^{\BF_1}_{(-1,1)} (t_1, t_2; \vec z; (1/2,-1/2)) = 1~.
\end{array}
\eea}
Hence, the Hilbert series for $SU(2)$ instantons on $\BF_1$ with $\hat{c}=1, \; \hat{k}=1/4$ is
\bea \label{HchalfSU2}
H[\hat{k}=1/4; SU(2); \hat{c}=1; \BF_1] (t_1,t_2,\vec z) = \frac{1}{1-z_2 z_1^{-1}} + \frac{1}{1-z_1 z_2^{-1}}  =1~.
\eea
The moduli space is a point, \ie~ zero complex dimensional.  For the King quiver \eref{quivSUN}, this corresponds to $k_1=0, \; k_2=1, \; N=2$.

\subsubsection*{The instanton number $\hat{k} =5/4$}
The contributions to $q^{5/4}$ in \eref{blowup} come from the product of the $\ell$-factors in \eref{Lfacc1} and the Hilbert series of 1 instanton on $\BC^2$:
\bea \label{HcthreehalfSU2}
&H[\hat{k} =5/4; SU(2); \hat{c}=1; \BF_1] (t_1,t_2,\vec z)  \nn \\
&= \frac{1}{1-z_2 z_1^{-1}}  \left[ \frac{\left(t_1+1\right) t_2 z_1 z_2}{\left(\frac{t_1}{t_2}-1\right) \left(t_2-1\right) \left(t_1 t_2 z_1-z_2\right) \left(t_1 z_2-t_2 z_1\right)}  + (t_1 \leftrightarrow t_2) \right] \nn \\
& \quad+  (z_1 \leftrightarrow z_2)  \nn \\
&= \PE[ t_1 + t_2] \times \PE[t^2(1+ z_1 z_2^{-1}+z_2 z_1^{-1}) - t^4]  \nn \\
&=  H[k=1; SU(2); \BC^2] (t_1, t_2, \vec z)~,
\eea
where $H[k=1; SU(2); \BC^2] (t_1, t_2, \vec z)$ denotes the Hilbert series \eref{HS1SU2C2} of one $SU(2)$ instanton on $\BC^2$; this is equal to the product of the Hilbert series of $\BC^2 \times (\BC^2/\BZ_2)$. Note that the moduli space is $4$ complex dimensional.

For the King quiver \eref{quivSUN}, this configuration corresponds to $k_1=1, \; k_2=2, \; N=2$.

\subsection{$USp(4)$ instantons}
In this subsection, we apply the blowup formula \eref{blowup} to compute the Hilbert series of $USp(4)$ instantons.  For the generating function $\CH[USp(4); \BF_1](t_1,t_2; \vec z; q)$ up to $q^{2}$, the relevants elements of the set of fixed points $\CV_{USp(4)}$ are 
\bea
\vec v  &= (-\sqrt{2}, -\sqrt{2}), \quad (-\sqrt{2}, 0), \quad (-\sqrt{2}, \sqrt{2}), \quad (0, -\sqrt{2}), \nn \\
&\qquad (0, 0), \quad (0, \sqrt{2}), \quad (\sqrt{2}, -\sqrt{2}),\quad  (\sqrt{2}, 0), \quad (\sqrt{2}, \sqrt{2})~.
\eea 
The relevant products of the $\ell$-factors are
{\small
\bea \label{USp4Lfactors}
\prod_{\vec \alpha \in C_2} \ell_{\vec \alpha}^{\BF_1} (t_1,t_2; \vec z; (0,0)) &=1~,  \\
\prod_{\vec \alpha \in C_2} \ell_{\vec \alpha}^{\BF_1} (t_1,t_2; \vec z; (0,\sqrt{2})) &=\left(1-z_1^{-\frac{1}{\sqrt{2}}} z_2^{-\frac{1}{\sqrt{2}}}\right) \left(1-z_1^{\frac{1}{\sqrt{2}}} z_2^{-\frac{1}{\sqrt{2}}}\right) \left(1-z_2^{-\sqrt{2}}\right) \times \nn \\
& \quad  \left(1-t_1^{-1} {z_2^{-\sqrt{2}}}\right) \left(1-t_2^{-1} {z_2^{-\sqrt{2}}}\right) \left(1-t_1 t_2 z_2^{\sqrt{2}}\right)~, \nn \\
\prod_{\vec \alpha \in C_2} \ell_{\vec \alpha}^{\BF_1} (t_1,t_2; \vec z; (\sqrt{2},\sqrt{2}))  &= \left(1-z_1^{-\sqrt{2}}\right) \left(1-z_1^{-\frac{1}{\sqrt{2}}} z_2^{-\frac{1}{\sqrt{2}}}\right) \left(1-z_2^{-\sqrt{2}}\right) \left(1-t_1^{-1} {z_1^{-\sqrt{2}}}\right) \times \nn \\
& \quad  \left(1-t_2^{-1} {z_1^{-\sqrt{2}}}\right)  \left(1-t_1 t_2 z_1^{\sqrt{2}}\right) \left(1-t_1^{-1} {z_1^{-\frac{1}{\sqrt{2}}} z_2^{-\frac{1}{\sqrt{2}}}}\right) \left(1-t_2^{-1} {z_1^{-\frac{1}{\sqrt{2}}} z_2^{-\frac{1}{\sqrt{2}}}}\right) \times  \nn \\
& \quad \left(1-t_1 t_2 z_1^{\frac{1}{\sqrt{2}}} z_2^{\frac{1}{\sqrt{2}}}\right) \left(1-t_1^{-1} {z_2^{-\sqrt{2}}}\right) \left(1-t_2^{-1} {z_2^{-\sqrt{2}}}\right) \left(1-t_1 t_2 z_2^{\sqrt{2}}\right)~,\nn
\eea}
with 
\bea \label{identitiesL}
\ell_{\vec \alpha}^{\BF_1} (t_1,t_2; (z_1,z_2); (v_1,v_2)) &= \ell_{\vec \alpha}^{\BF_1} (t_1,t_2; (z_2,z_1); (v_2,v_1))~,  \nn \\
\ell_{\vec \alpha}^{\BF_1} (t_1,t_2; (z_1,z_2); (v_1,-v_2)) &= \ell_{\vec \alpha}^{\BF_1} (t_1,t_2; (z_1,z_2^{-1}); (v_1,v_2))~,  \nn \\
\ell_{\vec \alpha}^{\BF_1} (t_1,t_2; (z_1,z_2); (-v_1,v_2)) &= \ell_{\vec \alpha}^{\BF_1} (t_1,t_2; (z_1^{-1},z_2); (v_1,v_2))~.
\eea
The blowup formula \eref{blowup} gives
\bea
\CH[USp(4); \BF_1](t_1,t_2; \vec z; q) 
&= 1+ H[1; USp(4); \BC^2](t_1, t_2; \vec z^{1/\sqrt{2}}) \; q \nn\\
& \qquad + H[2; USp(4); \BC^2](t_1, t_2; \vec z^{1/\sqrt{2}}) \; q^2 + \ldots~. 
\eea

\subsection{$SO(5)$ instantons}
In this subsection, we apply the blowup formula \eref{blowup} to compute the Hilbert series of $SO(5)$ instantons.  

\subsubsection{The case of $\hat{c}=0$}
For $\hat{c}=0$, the generating function $\CH[SO(5); \BF_1](t_1,t_2; \vec z; q)$ up to $q^{2}$ arises from the following elements of the set of fixed points $\CV_{SO(5)}(\hat{c}=0)$ are 
\bea
\vec v  &= (-2, 0), \quad (-1, -1), \quad (-1, 1), \quad (0, -2), \quad (0, 0), \nn \\
& \qquad (0, 2), \quad (1, -1), \quad (1, 1), \quad (2, 0)~.
\eea 
The fugacities $\vec z=(z_1,z_2)$ for $USp(4)$ in the previous subsection are related to those of $\vec x =(x_1, x_2)$ for $SO(5)$ by
\bea
x_1 = (z_1 z_2)^{1/\sqrt{2}}~, \qquad x_2 = (z_1 z_2^{-1})^{1/\sqrt{2}}~.
\eea
The relevant $\ell$-factors are similar to \eref{USp4Lfactors}; for example, 
\bea
\prod_{\vec \alpha \in B_2} \ell_{\vec \alpha}^{\BF_1} (t_1,t_2; \vec x; (0,2)) =\prod_{\vec \alpha \in C_2}  \ell_{\vec \alpha}^{\BF_1} (t_1,t_2; \vec z; (\sqrt{2}, -\sqrt{2}))~.
\eea
Using the blowup formula \eref{blowup}, we find that
\bea \label{SO5blowupc10}
\CH[SO(5); \hat{c}=0; \BF_1](t_1,t_2; \vec x; q) 
&= 1+ H[1; SO(5); \BC^2](t_1, t_2; \vec x) \; q \nn\\
& \qquad + H[2; SO(5); \BC^2](t_1, t_2; \vec x) \; q^2 + \ldots~. 
\eea
This result is in agreement with the Hilbert series computed from the King construction \eref{ADHMSO}, with $N=5$ and $k$ being integer-valued equal to the powers of $q$ in \eref{SO5blowupc10}.  

\subsubsection{The case of $\hat{c}=1$}
Up to order $q^{3/2}$ of the generating function $\CH[SO(5); \hat{c}=1; \BF_1](t_1,t_2; \vec x; q)$ given by \eref{blowup}, the relevant elements of the set of fixed points $\CV_{SO(5)}(\hat{c}=1)$ are 
\bea
\vec v = (1, 0),  \quad (0, 1), \quad (-1, 0), \quad (0, -1)~.
\eea
The relevant products of $\ell$-factors are given by
\bea
\prod_{\vec \alpha \in B_2} \ell_{\vec \alpha}^{\BF_1} (t_1,t_2; \vec x; (1,0)) = \left(1-\frac{1}{x_1}\right) \left(1-\frac{1}{x_1 x_2}\right) \left(1-\frac{x_2}{x_1}\right)~,
\eea
together with \eref{identitiesL}.  Using the blowup formula \eref{blowup}, we find that
\bea \label{SO5blowupc11}
& \CH[SO(5); \hat{c}=1; \BF_1](t_1,t_2; \vec x; q)  \nn \\
&= q^{1/2}+ H[1; SO(5); \BC^2](t_1, t_2; \vec x) \; q^{3/2}  + \ldots~. 
\eea
This result is in agreement with the Hilbert series computed from the King construction \eref{ADHMSO}, with $N=5$ and $k$ being half-odd-integer-valued equal to the powers of $q$ in \eref{SO5blowupc11}.

\subsection{$SO(6)$ instantons}
In this subsection, we apply the blowup formula \eref{blowup} to compute the Hilbert series of $SO(6)$ instantons.  

\subsubsection{The case of $\hat{c}=0$}
For $\hat{c}=0$, the generating function $\CH[SO(6); \BF_1](t_1,t_2; \vec z; q)$ up to $q^{2}$ arises from the following elements of the set of fixed points $\CV_{SO(6)}(\hat{c}=0)$ are 
\bea
& (-2,0,0), \quad (-1,-1,0), \quad (-1,0,-1), \quad (-1,0,1), \quad (-1,1,0), \quad (0,-2,0), \nn \\
& (0,-1,-1), \quad (0,-1,1), \quad (0,0,-2), \quad (0,0,0), \quad (0,0,2), \quad (0,1,-1), \quad (0,1,1), \nn \\
& (0,2,0), \quad (1,-1,0), \quad (1,0,-1), \quad (1,0,1), \quad (1,1,0), \quad (2,0,0)~.
\eea 
Using the blowup formula \eref{blowup}, we find that
\bea \label{SO6blowupc10}
\CH[SO(6); \hat{c}=0; \BF_1](t_1,t_2; \vec x; q) 
&= 1+ H[1; SO(6); \BC^2](t_1, t_2; \vec x) \; q \nn\\
& \qquad + H[2; SO(6); \BC^2](t_1, t_2; \vec x) \; q^2 + \ldots~. 
\eea
This result is in agreement with the Hilbert series computed from the King construction \eref{ADHMSO}, with $N=6$ and $k$ being integer-valued equal to the powers of $q$ in \eref{SO6blowupc10}.  

\subsubsection{The case of $\hat{c}=1$}
Up to order $q^{5/2}$ of the generating function $\CH[SO(6); \hat{c}=1; \BF_1](t_1,t_2; \vec x; q)$ given by \eref{blowup}, the relevant elements of the set of fixed points $\CV_{SO(6)}(\hat{c}=1)$ are 
\bea
&(-1,-1,-1), \quad (-1,-1,1), \quad (-1,0,0), \quad (-1,1,-1), \quad (-1,1,1),  \\
&(0,-1,0), \quad (0,0,-1), \quad (0,0,1), \quad (0,1,0), \quad (1,-1,-1), \quad (1,-1,1), \nn \\
& (1,0,0), \quad (1,1,-1), \quad (1,1,1)~.\nn
\eea
Using the blowup formula \eref{blowup}, we find that
\bea \label{SO6blowupc11}
& \CH[SO(6); \hat{c}=1; \BF_1](t_1,t_2; \vec x; q)  \nn \\
&= q^{1/2}+ H[1; SO(6); \BC^2](t_1, t_2; \vec x) \; q^{3/2}+ H[2; SO(6); \BC^2](t_1, t_2; \vec x) \; q^{5/2}  + \ldots~. 
\eea
This result is in agreement with the Hilbert series computed from the King construction \eref{ADHMSO}, with $N=6$ and $k$ being half-odd-integer-valued equal to the powers of $q$ in \eref{SO6blowupc11}.

\subsection{$G_2$ instantons}
Let us apply the blowup formula to compute the Hilbert series of $G_2$ instantons on $\CP^2$.  In this case, there is no known construction from a quiver gauge theory.  However, it is still possible to compute the Hilbert series using the blowup formula in a similar way to \cite{Keller:2012da} as we demonstrate below.

Up to order $q$ of the generating function $\CH[G_2; \BF_1](t_1,t_2; \vec x; q)$ given by \eref{blowup}, the relevant elements of the set of fixed points $\CV_{G_2}$ are
\bea
& (0,0), \quad \left(0,\sqrt{2}\right ), \quad \left(\sqrt{\frac{3}{2}},\frac{1}{\sqrt{2}}\right ), \quad \left(-\sqrt{\frac{3}{2}},\frac{1}{\sqrt{2}}\right ), \nn \\
& \left(\sqrt{\frac{3}{2}},-\frac{1}{\sqrt{2}}\right ), \quad \left(0,-\sqrt{2}\right ), \quad \left(-\sqrt{\frac{3}{2}},-\frac{1}{\sqrt{2}}\right )~.
\eea
The relevants products of $\ell$-factors are given by
\bea
&\prod_{\vec \alpha \in G_2} \ell_{\vec \alpha}^{\BF_1} (t_1,t_2; \vec x; (0,0)) = 1~, \nn \\ 
& \prod_{\vec \alpha \in G_2} \ell_{\vec \alpha}^{\BF_1} (t_1,t_2; \vec x; (0,\sqrt{2})) = \left(1-z_1^{-\sqrt{\frac{3}{2}}} z_2^{-\frac{1}{\sqrt{2}}}\right) \left(1-z_1^{\sqrt{\frac{3}{2}}} z_2^{-\frac{1}{\sqrt{2}}}\right) \left(1-z_1^{-\frac{1}{\sqrt{6}}} z_2^{-\frac{1}{\sqrt{2}}}\right) \times  \nn \\
& \qquad \qquad \left(1-z_1^{\frac{1}{\sqrt{6}}} z_2^{-\frac{1}{\sqrt{2}}}\right) \left(1-z_2^{-\sqrt{2}}\right) \left(1-\frac{z_2^{-\sqrt{2}}}{t_1}\right) \left(1-\frac{z_2^{-\sqrt{2}}}{t_2}\right) \left(1-t_1 t_2 z_2^{\sqrt{2}}\right)~, \nn \\
&\prod_{\vec \alpha \in G_2} \ell_{\vec \alpha}^{\BF_1} \left(t_1,t_2; \vec x; \left (\sqrt{\frac{3}{2}},\frac{1}{\sqrt{2}}\right ) \right) = \left(1-z_1^{-\sqrt{\frac{2}{3}}}\right) \left(1-z_1^{-\sqrt{\frac{3}{2}}} z_2^{-\frac{1}{\sqrt{2}}}\right) \left(1-z_1^{-\frac{1}{\sqrt{6}}} z_2^{-\frac{1}{\sqrt{2}}}\right) \nn \\
& \qquad \qquad \left(1-z_1^{-\sqrt{\frac{3}{2}}} z_2^{\frac{1}{\sqrt{2}}}\right) \left(1-z_2^{-\sqrt{2}}\right) \left(1-\frac{z_1^{-\sqrt{\frac{3}{2}}} z_2^{-\frac{1}{\sqrt{2}}}}{t_1}\right) \left(1-\frac{z_1^{-\sqrt{\frac{3}{2}}} z_2^{-\frac{1}{\sqrt{2}}}}{t_2}\right) \times \nn \\
&\qquad \qquad \left(1-t_1 t_2 z_1^{\sqrt{\frac{3}{2}}} z_2^{\frac{1}{\sqrt{2}}}\right)~,
\eea
together with the identities \eref{identitiesL}.  Using the blowup formula \eref{blowup}, we find that
\bea \label{G2blowupc10}
\CH[G_2; \BF_1](t_1,t_2; \vec x; q) = 1+ H[1; G_2; \BC^2](t_1, t_2; \vec x) \; q +\ldots~.
\eea

\section{Conclusions}\label{conclusions}

In this paper we have studied self-dual connections on $\CP^2$. These are described by an ADHM-like connection introduced in the mathematical literature in \cite{buchdal1,buchdal2,buchdal3,king, bryansanders2}. We have embedded such construction into a supersymmetric gauge theory. Surprisingly, and as opposed to the well-known ADHM constructions for instantons on $S^4$, in the case at hand such gauge theory is naturally a $3d$ gauge theory with 4 supercharges. The low amount of supersymmetry allows for quantum corrections that render large anomalous dimensions of the chiral fields at the conformal fixed point. These are crucial to correctly obtain, in the suitable instanton branch, the expected properties for the moduli space of instantons of a pure gauge theory, such as the hyperK\"ahler structure. However, it is important to stress that quantum effects due to monopole operators in the instanton branch vanish.

Since $\CP^2$ is a topologically non-trivial space, the structure of the instanton sector is more involved than in the $S^4$ counterpart. In particular, the global structure of the gauge group becomes manifest. As we have argued, in the case of unitary instantons, we are led to conclude that the gauge group is $SU(N)/\mathbb{Z}_N$. Because of this, the first Chern class of the \textit{a priori} $U(N)$ instanton gets reduced mod $N$. We have encountered a similar phenomenon for orthogonal instantons, only that in this case mod 2. On the other hand, symplectic instantons don't seem to exhibit such phenomenon. In fact, it would be interesting to reconsider in deeper detail this point, as well as the exact global properties of the gauge group for orthogonal and unitary instantons along the lines of \cite{Aharony:2013hda}.

The Hilbert series is only sensitive to the mesonic type operators in the in the instanton branch $\mathcal{M}^{G}_{\CP^2}$. However there is a resolved moduli space $\widehat{\mathcal{M}}^{G}_{\CP^2}$, which is a non-singular space with a generically larger dimension than the former: ${\rm dim}_\BC\widehat{\mathcal{M}}^{G}_{\CP^2} \geq \dim_\BC\mathcal{M}^{G}_{\CP^2}$.\footnote{In a recent work \cite{Tachikawa:2014qaa}, the moduli space of $SO(8)$ instantons on a smooth ALE space has been studied from the perspective of the worldvolume theory M5-branes on Riemann surfaces.  The hyperK\"ahler quotient of such a moduli space is described.  Indeed, the dimension of the moduli space of $SO(8)$ instanton on a smooth ALE space is found to be larger than the dimension of its counterpart on the singular orbifold (see also \cite{Intriligator:1997kq}).} It is natural to ask to classify operators that parametrise such `extra directions'. These might involve baryon-like operators, Wilson lines or monopole operators, possibly slightly departing from the strict definition of the instanton branch we have used where all monopole operators are set to zero. It would be very interesting to study such an aspect of the resolved space. Note that the parameter $\hat{c}$, which has an interpretation as the first Chern class for unitary instantons, can be regarded as a  dissolved surface operator, since, at least morally speaking, $\hat{c}\sim\int_{\CP^1\subset \CP^2}{\rm Tr}\,F$. In fact, we can consider a special case in which either $k_1$ or $k_2$ is zero; then, the quiver \eref{quivSUN} becomes a quiver theory with a rank $\hat{c}$ gauge group, with $0\leq \hat{c}\leq N-1$. Naively, for the appropriate FI choice, we expect the latter to describe a moduli space of vortices, which, being typically a compact space, cannot be parametrised in terms of holomorphic gauge invariant quantities as for the case of non-compact hyperK\"ahler cone.  The study of such vortex moduli space might provides deeper insights into the `extra directions' described above in the context of the resolved moduli space  (of course, similar observations can be made for orthogonal instantons). It will be very interesting to pursue this further, perhaps along the lines of \cite{Hanany:2014hia}, task which we leave for future research.

At least for the unitary case, the gauge theory in which we have embedded the ADHM construction of $\CP^2$ instantons admits a stringy interpretation as the worldvolume theory on M2 branes probing a Calabi-Yau 4-fold singularity \cite{Benini:2009qs}. This opens up the possibility of an explicit geometric study of the instanton properties. In particular, holographic tools along the lines of \cite{Bergman:2012qh,Dey:2013fea} could be of great interest. Of course, it would be interesting to extend this analysis and explore the precise stringy realization of the orthogonal and symplectic cases. We will postpone these studies for future work.

As mentioned above, our ADHM instantons have the same duality properties as the K\"ahler form, and hence naively do not seem to be the ones relevant for localization of gauge theories on $\CP^2$. However, it would be interesting to clarify wether they might contribute under some circumstances to partition functions. On a related note, the blowup formula localizes on two contributions. It would be interesting to clarify possible relations to the factorization properties of instanton partition functions recently discussed in \textit{e.g.} \cite{Nieri:2013yra,Nieri:2013vba,Qiu:2014oqa}.

\section*{Acknowledgements}

We would like to thank S. Benvenuti, S. Cremonesi, S. Franco, A. Hanany, K. Lee, P. Meessen, B. Pioline, J. Schmude, J. Song and H. Triendl for useful conversations. We would like to specially thank A. King for providing us with a copy of \cite{king} as well as for correspondences. The work of D.~R-G. is partially supported by the Ramon y Cajal fellowship RyC-2011-07593, as well as by the Spanish government grant FPA2012-35043-C0-02 and the EU CIG grant UE-14-GT5LD2013-618459.  N.~M.~ would like to express his gratitude to the String Theory Group of the University of Oviedo; the program `New Methods in Nonperturbative Quantum Field Theory' at Kavli Institute for Theoretical Physics (KITP); W. Yan and the Caltech Particle Theory Group; A. Neitzke, A. Dey, J. Distler and the University of Texas at Austin for their very kind hospitality and support.   We would like to acknowledge the support of the COST action MP1210 STSM.  This research was also supported in part by the National Science Foundation under Grant No. NSF PHY11-25915. 

\bibliographystyle{ytphys}
\bibliography{ref}

\providecommand{\href}[2]{#2}\begingroup\raggedright\begin{thebibliography}{10}

\bibitem{buchdal1}
N.~Buchdal, ``{Instantons on CP2},'' {\em J.Diff.Geom.} {\bfseries 24} (1986)
  19--52.

\bibitem{buchdal2}
N.~Buchdal, ``{Stable 2-bundles on HIrzebruch surfaces},'' {\em Math. Z.}
  {\bfseries 194} (1987) 143--152.

\bibitem{buchdal3}
N.~Buchdal, ``{Hermitiean-Einstein Connections and Stable Vector Bundles Over
  Compact Algebraic Surfaces},'' {\em Math. Ann.} {\bfseries 280} (1988)
  625--648.

\bibitem{king}
A.~King, {\em Instantons and holomorphic bundles on the blown-up plane}.
\newblock PhD thesis, Thesis, Oxford University, 1989.

\bibitem{bryansanders2}
J.~Bryan and M.~Sanders, ``Instantons on \{S4\} and cp2, rank stabilization,
  and bott periodicity,''
  \href{http://www.sciencedirect.com/science/article/pii/S0040938399000191}{{\em
  Topology} {\bfseries 39} no.~2, (2000) 331 -- 352}.

\bibitem{ADHM}
M.~Atiyah, V.~Drinfeld, N.~Hitchin, and Y.~Manin, ``{Construction of
  instantons},'' {\em Physics Letters A} {\bfseries 65 (3)} 185--187.

\bibitem{Atiyah:1979iu}
M.~Atiyah,
``{GEOMETRY OF YANG-MILLS FIELDS},''.

\bibitem{DonaldsonKronheimer}
{S.Donaldson and P.Kronheimer }, {\em The geometry of four-manifolds}.
\newblock Oxford University Press, 1990.

\bibitem{Aharony:2013hda}
O.~Aharony, N.~Seiberg, and Y.~Tachikawa, ``{Reading between the lines of
  four-dimensional gauge theories},''
  \href{http://dx.doi.org/10.1007/JHEP08(2013)115}{{\em JHEP} {\bfseries 1308}
  (2013) 115},
\href{http://arxiv.org/abs/1305.0318}{{\ttfamily arXiv:1305.0318}}.

\bibitem{Kim:2012qf}
H.-C. Kim, J.~Kim, and S.~Kim, ``{Instantons on the 5-sphere and M5-branes},''
\href{http://arxiv.org/abs/1211.0144}{{\ttfamily arXiv:1211.0144 [hep-th]}}.

\bibitem{Douglas:1995bn}
M.~R. Douglas, ``{Branes within branes},''
\href{http://arxiv.org/abs/hep-th/9512077}{{\ttfamily arXiv:hep-th/9512077
  [hep-th]}}.

\bibitem{Witten:1995gx}
E.~Witten, ``{Small instantons in string theory},''
  \href{http://dx.doi.org/10.1016/0550-3213(95)00625-7}{{\em Nucl.Phys.}
  {\bfseries B460} (1996) 541--559},
\href{http://arxiv.org/abs/hep-th/9511030}{{\ttfamily arXiv:hep-th/9511030
  [hep-th]}}.

\bibitem{Douglas:1996sw}
M.~R. Douglas and G.~W. Moore, ``{D-branes, quivers, and ALE instantons},''
\href{http://arxiv.org/abs/hep-th/9603167}{{\ttfamily arXiv:hep-th/9603167
  [hep-th]}}.

\bibitem{Benvenuti:2010pq}
S.~Benvenuti, A.~Hanany, and N.~Mekareeya, ``{The Hilbert Series of the One
  Instanton Moduli Space},''
  \href{http://dx.doi.org/10.1007/JHEP06(2010)100}{{\em JHEP} {\bfseries 1006}
  (2010) 100},
\href{http://arxiv.org/abs/1005.3026}{{\ttfamily arXiv:1005.3026 [hep-th]}}.

\bibitem{Hanany:2012dm}
A.~Hanany, N.~Mekareeya, and S.~S. Razamat, ``{Hilbert Series for Moduli Spaces
  of Two Instantons},'' \href{http://dx.doi.org/10.1007/JHEP01(2013)070}{{\em
  JHEP} {\bfseries 1301} (2013) 070},
\href{http://arxiv.org/abs/1205.4741}{{\ttfamily arXiv:1205.4741 [hep-th]}}.

\bibitem{Vafa:1994tf}
C.~Vafa and E.~Witten, ``{A Strong coupling test of S duality},''
  \href{http://dx.doi.org/10.1016/0550-3213(94)90097-3}{{\em Nucl.Phys.}
  {\bfseries B431} (1994) 3--77},
\href{http://arxiv.org/abs/hep-th/9408074}{{\ttfamily arXiv:hep-th/9408074
  [hep-th]}}.

\bibitem{Fucito:2004ry}
F.~Fucito, J.~F. Morales, and R.~Poghossian, ``{Multi instanton calculus on ALE
  spaces},'' \href{http://dx.doi.org/10.1016/j.nuclphysb.2004.09.014}{{\em
  Nucl.Phys.} {\bfseries B703} (2004) 518--536},
\href{http://arxiv.org/abs/hep-th/0406243}{{\ttfamily arXiv:hep-th/0406243
  [hep-th]}}.

\bibitem{Bonelli:2011kv}
G.~Bonelli, K.~Maruyoshi, and A.~Tanzini, ``{Gauge Theories on ALE Space and
  Super Liouville Correlation Functions},''
  \href{http://dx.doi.org/10.1007/s11005-012-0553-x}{{\em Lett.Math.Phys.}
  {\bfseries 101} (2012) 103--124},
\href{http://arxiv.org/abs/1107.4609}{{\ttfamily arXiv:1107.4609 [hep-th]}}.

\bibitem{Dey:2013fea}
A.~Dey, A.~Hanany, N.~Mekareeya, D.~Rodr'guez-G—mez, and R.-K. Seong,
  ``{Hilbert Series for Moduli Spaces of Instantons on
  $\mathbb{C}$$^{2}$/$\mathbb{Z}$$_{n}$},''
  \href{http://dx.doi.org/10.1007/JHEP01(2014)182}{{\em JHEP} {\bfseries 1401}
  (2014) 182},
\href{http://arxiv.org/abs/1309.0812}{{\ttfamily arXiv:1309.0812 [hep-th]}}.

\bibitem{Ito:2013kpa}
Y.~Ito, K.~Maruyoshi, and T.~Okuda, ``{Scheme dependence of instanton counting
  in ALE spaces},'' \href{http://dx.doi.org/10.1007/JHEP05(2013)045}{{\em JHEP}
  {\bfseries 1305} (2013) 045},
\href{http://arxiv.org/abs/1303.5765}{{\ttfamily arXiv:1303.5765 [hep-th]}}.

\bibitem{Fucito:2006kn}
F.~Fucito, J.~F. Morales, and R.~Poghossian, ``{Instanton on toric
  singularities and black hole countings},''
  \href{http://dx.doi.org/10.1088/1126-6708/2006/12/073}{{\em JHEP} {\bfseries
  0612} (2006) 073},
\href{http://arxiv.org/abs/hep-th/0610154}{{\ttfamily arXiv:hep-th/0610154
  [hep-th]}}.

\bibitem{Gasparim:2008ri}
E.~Gasparim and C.-C.~M. Liu, ``{The Nekrasov Conjecture for Toric Surfaces},''
  \href{http://dx.doi.org/10.1007/s00220-009-0948-4}{{\em Commun.Math.Phys.}
  {\bfseries 293} (2010) 661--700},
\href{http://arxiv.org/abs/0808.0884}{{\ttfamily arXiv:0808.0884 [math.AG]}}.

\bibitem{Bruzzo:2009uc}
U.~Bruzzo, R.~Poghossian, and A.~Tanzini, ``{Poincare polynomial of moduli
  spaces of framed sheaves on (stacky) Hirzebruch surfaces},''
  \href{http://dx.doi.org/10.1007/s00220-011-1231-z}{{\em Commun.Math.Phys.}
  {\bfseries 304} (2011) 395--409},
\href{http://arxiv.org/abs/0909.1458}{{\ttfamily arXiv:0909.1458 [math.AG]}}.

\bibitem{Bonelli:2012ny}
G.~Bonelli, K.~Maruyoshi, A.~Tanzini, and F.~Yagi, ``{N=2 gauge theories on
  toric singularities, blow-up formulae and W-algebrae},''
  \href{http://dx.doi.org/10.1007/JHEP01(2013)014}{{\em JHEP} {\bfseries 1301}
  (2013) 014},
\href{http://arxiv.org/abs/1208.0790}{{\ttfamily arXiv:1208.0790 [hep-th]}}.

\bibitem{Benini:2009qs}
F.~Benini, C.~Closset, and S.~Cremonesi, ``{Chiral flavors and M2-branes at
  toric CY4 singularities},''
  \href{http://dx.doi.org/10.1007/JHEP02(2010)036}{{\em JHEP} {\bfseries 1002}
  (2010) 036},
\href{http://arxiv.org/abs/0911.4127}{{\ttfamily arXiv:0911.4127 [hep-th]}}.

\bibitem{Nakajima:2008eq}
H.~Nakajima and K.~Yoshioka, ``{Perverse coherent sheaves on blow-up. I. A
  Quiver description},''
\href{http://arxiv.org/abs/0802.3120}{{\ttfamily arXiv:0802.3120 [math.AG]}}.

\bibitem{Nakajima:2005fg}
H.~Nakajima and K.~Yoshioka, ``{Instanton counting on blowup. II. K-theoretic
  partition function},''
\href{http://arxiv.org/abs/math/0505553}{{\ttfamily arXiv:math/0505553
  [math-ag]}}.

\bibitem{Keller:2012da}
C.~A. Keller and J.~Song, ``{Counting Exceptional Instantons},''
  \href{http://dx.doi.org/10.1007/JHEP07(2012)085}{{\em JHEP} {\bfseries 1207}
  (2012) 085},
\href{http://arxiv.org/abs/1205.4722}{{\ttfamily arXiv:1205.4722 [hep-th]}}.

\bibitem{Donaldson:1984tm}
S.~Donaldson, ``{INSTANTONS AND GEOMETRIC INVARIANT THEORY},''
\href{http://dx.doi.org/10.1007/BF01212289}{{\em Commun.Math.Phys.} {\bfseries
  93} (1984) 453--460}.

\bibitem{Donaldson:1985zz}
S.~Donaldson, ``{ANTI SELF-DUAL YANG-MILLS CONNECTIONS OVER COMPLEX ALGEBRAIC
  SURFACES AND STABLE VECTOR BUNDLES},''
\href{http://dx.doi.org/10.1112/plms/s3-50.1.1}{{\em Proc.Lond.Math.Soc.}
  {\bfseries 50} (1985) 1--26}.

\bibitem{Benini:2007kg}
F.~Benini, ``{A Chiral cascade via backreacting D7-branes with flux},''
  \href{http://dx.doi.org/10.1088/1126-6708/2008/10/051}{{\em JHEP} {\bfseries
  0810} (2008) 051},
\href{http://arxiv.org/abs/0710.0374}{{\ttfamily arXiv:0710.0374 [hep-th]}}.

\bibitem{Davey:2009sr}
J.~Davey, A.~Hanany, N.~Mekareeya, and G.~Torri, ``{Phases of M2-brane
  Theories},'' \href{http://dx.doi.org/10.1088/1126-6708/2009/06/025}{{\em
  JHEP} {\bfseries 0906} (2009) 025},
\href{http://arxiv.org/abs/0903.3234}{{\ttfamily arXiv:0903.3234 [hep-th]}}.

\bibitem{M2}
D.~R. Grayson and M.~E. Stillman, ``Macaulay2, a software system for research
  in algebraic geometry.'' Available at
  \href{http://www.math.uiuc.edu/Macaulay2/}%
  {http://www.math.uiuc.edu/Macaulay2/}.

\bibitem{Witten:2000nv}
E.~Witten, ``{Supersymmetric index in four-dimensional gauge theories},'' {\em
  Adv.Theor.Math.Phys.} {\bfseries 5} (2002) 841--907,
\href{http://arxiv.org/abs/hep-th/0006010}{{\ttfamily arXiv:hep-th/0006010
  [hep-th]}}.

\bibitem{bryansanders1}
J.~Bryan and M.~Sanders, ``The rank stable topology of instantons of
  {$\overline{\bold C{\rm P}}{}^2$},''
  \href{http://dx.doi.org/10.1090/S0002-9939-97-04156-7}{{\em Proc. Amer. Math.
  Soc.} {\bfseries 125} no.~12, (1997) 3763--3768}.

\bibitem{Nakajima:2003pg}
H.~Nakajima and K.~Yoshioka, ``{Instanton counting on blowup. 1.},''
  \href{http://dx.doi.org/10.1007/s00222-005-0444-1}{{\em Invent.Math.}
  {\bfseries 162} (2005) 313--355},
\href{http://arxiv.org/abs/math/0306198}{{\ttfamily arXiv:math/0306198
  [math-ag]}}.

\bibitem{Flume:2002az}
R.~Flume and R.~Poghossian, ``{An Algorithm for the microscopic evaluation of
  the coefficients of the Seiberg-Witten prepotential},''
  \href{http://dx.doi.org/10.1142/S0217751X03013685}{{\em Int.J.Mod.Phys.}
  {\bfseries A18} (2003) 2541},
\href{http://arxiv.org/abs/hep-th/0208176}{{\ttfamily arXiv:hep-th/0208176
  [hep-th]}}.

\bibitem{Nekrasov:2002qd}
N.~A. Nekrasov, ``{Seiberg-Witten prepotential from instanton counting},''
  \href{http://dx.doi.org/10.4310/ATMP.2003.v7.n5.a4}{{\em
  Adv.Theor.Math.Phys.} {\bfseries 7} (2004) 831--864},
\href{http://arxiv.org/abs/hep-th/0206161}{{\ttfamily arXiv:hep-th/0206161
  [hep-th]}}.

\bibitem{Bruzzo:2002xf}
U.~Bruzzo, F.~Fucito, J.~F. Morales, and A.~Tanzini, ``{Multiinstanton calculus
  and equivariant cohomology},''
  \href{http://dx.doi.org/10.1088/1126-6708/2003/05/054}{{\em JHEP} {\bfseries
  0305} (2003) 054},
\href{http://arxiv.org/abs/hep-th/0211108}{{\ttfamily arXiv:hep-th/0211108
  [hep-th]}}.

\bibitem{Tachikawa:2014qaa}
Y.~Tachikawa, ``{Moduli spaces of SO(8) instantons on smooth ALE spaces as
  Higgs branches of 4d N=2 supersymmetric theories},''
\href{http://arxiv.org/abs/1402.4200}{{\ttfamily arXiv:1402.4200 [hep-th]}}.

\bibitem{Intriligator:1997kq}
K.~A. Intriligator, ``{RG fixed points in six-dimensions via branes at orbifold
  singularities},'' \href{http://dx.doi.org/10.1016/S0550-3213(97)00236-8}{{\em
  Nucl.Phys.} {\bfseries B496} (1997) 177--190},
\href{http://arxiv.org/abs/hep-th/9702038}{{\ttfamily arXiv:hep-th/9702038
  [hep-th]}}.

\bibitem{Hanany:2014hia}
A.~Hanany and R.-K. Seong, ``{Hilbert Series and Moduli Spaces of k U(N)
  Vortices},''
\href{http://arxiv.org/abs/1403.4950}{{\ttfamily arXiv:1403.4950 [hep-th]}}.

\bibitem{Bergman:2012qh}
O.~Bergman and D.~Rodriguez-Gomez, ``{Probing the Higgs branch of 5d fixed
  point theories with dual giant gravitons in AdS(6)},''
  \href{http://dx.doi.org/10.1007/JHEP12(2012)047}{{\em JHEP} {\bfseries 1212}
  (2012) 047},
\href{http://arxiv.org/abs/1210.0589}{{\ttfamily arXiv:1210.0589 [hep-th]}}.

\bibitem{Nieri:2013yra}
F.~Nieri, S.~Pasquetti, and F.~Passerini, ``{3d $\&$; 5d gauge theory partition
  functions as q-deformed CFT correlators},''
\href{http://arxiv.org/abs/1303.2626}{{\ttfamily arXiv:1303.2626 [hep-th]}}.

\bibitem{Nieri:2013vba}
F.~Nieri, S.~Pasquetti, F.~Passerini, and A.~Torrielli, ``{5D partition
  functions, q-Virasoro systems and integrable spin-chains},''
\href{http://arxiv.org/abs/1312.1294}{{\ttfamily arXiv:1312.1294 [hep-th]}}.

\bibitem{Qiu:2014oqa}
J.~Qiu, L.~Tizzano, J.~Winding, and M.~Zabzine, ``{Gluing Nekrasov partition
  functions},''
\href{http://arxiv.org/abs/1403.2945}{{\ttfamily arXiv:1403.2945 [hep-th]}}.

\end{thebibliography}\endgroup
\end{document}